\documentclass[10pt, twocolumn]{IEEEtran}
\usepackage{graphicx}
\usepackage{epsfig,amsfonts,amsmath,amssymb,amstext,amsthm}
\usepackage{threeparttable}
\usepackage{epstopdf}
\usepackage{footnote}
\usepackage{setspace}

\newcommand{\bm}[1]{\mbox{\pmb{$#1$}}}

\newcommand{\dref}[1]{(\ref{#1})}
\newcommand{\Tr}{\mbox{Tr}}
\newcommand{\perm}{\textup{perm}}
\newcommand{\Perm}{\textup{Perm}}

\newtheorem {Lemma}{Lemma}
\newtheorem {Corollary}{Corollary}

\begin{document}

\title{\Huge{Performance Analysis of Macrodiversity MIMO Systems with MMSE and ZF Receivers in Flat Rayleigh Fading}}

\author{Dushyantha A. Basnayaka,~\IEEEmembership{Student Member,~IEEE,}
        Peter J. Smith,~\IEEEmembership{Senior Member,~IEEE}
        and~Phillipa A. Martin,~\IEEEmembership{Senior Member,~IEEE}
\thanks{D. A. Basnayaka, P. J. Smith and P. A. Martin are with the Department of Electrical and Computer Engineering, University of Canterbury, Christchurch, New
Zealand. E-mail:\{dush, p.smith, p.martin\}@elec.canterbury.ac.nz.}
\thanks{D. A. Basnayaka is supported by a University of Canterbury International Doctoral Scholarship.}}

\maketitle

\begin{abstract}
Consider a multiuser system where an arbitrary number of users
communicate with a distributed receive array over independent
Rayleigh fading paths. The receive array performs minimum mean
squared error (MMSE) or zero forcing (ZF) combining and perfect
channel state information is assumed at the receiver. This scenario
is well-known and exact analysis is possible  when the receive
antennas are located in a single array. However, when the antennas
are distributed, the individual links all have different average
signal to noise ratio (SNRs) and this is a much more challenging
problem. In this paper, we provide approximate distributions for the
output SNR of a ZF receiver and the output signal to interference
plus noise ratio (SINR) of an MMSE receiver. In addition, simple
high SNR approximations are provided for the symbol error rate (SER)
of both receivers assuming $M$-PSK or $M$-QAM modulations. These
high SNR results provide array gain and diversity gain information
as well as a remarkably simple functional link between performance
and the link powers.
\end{abstract}


\section{Introduction}\label{sec:general:introduction}
With the advent of space diversity systems, decoupling users through
channel aware signal processing techniques in the presence of
multiple access interference (MAI) and noise has become an integral
part of the system design. There are various processing techniques
now widely adopted in research and standards \cite{Verdu98}. Among
them, linear combining methods are popular for their simplicity
despite the fact that they are not optimum in a maximum likelihood
sense. Two key linear combiners are zero forcing (ZF) and
minimum-mean squared-error (MMSE). Although they are not optimal,
the MMSE receiver satisfies an alternative criterion, i.e., it
minimizes the mean squared error (MSE) and ZF is known to eliminate
MAI completely.\\
The performance analysis of such linear receivers is of great
interest in wireless communication \cite{Winters84} as it provides a
baseline link level performance metric for the system. Today,
performance results for MMSE/ZF receivers are well known for
microdiversity systems where co-located diversity antennas at the
base station communicate with distributed users
\cite{Gao98,Winters94,Gore02}. Macro-scale diversity combining has
recently become more common from a variety of perspectives
\cite{Fosch06, Siva07}. Any system where both transmit and receive
antennas are widely separated can be interpreted as a macrodiversity
multiple input multiple output (MIMO) system. They occur naturally
in network MIMO systems \cite{Siva07,Sander09} and collaborative
MIMO concepts \cite[p.~69]{Big07} and \cite{SaKiMo10}.\\
The performance of macrodiversity systems has been investigated via
simulation \cite{Xiao03}, but very few analytical results appear to
be available. The reason for the lack of results is the complexity
of the channel matrix that arises in macrodiversity systems. When
the receive antennas are co-located, classical models and Kronecker
correlation matrix has a Wishart form. Here, extensive results in
multivariate statistics can be leveraged and performance analysis is
now well advanced. In contrast, the macrodiversity case violates the
Wishart assumptions and there is no such distribution in the
literature for macrodiversity channel matrices for finite size
systems. This makes most of the analytical work extremely difficult.
The analytical complexity is clearly evident even in the simplest
case of a dual source scenario \cite{Dush_mmse_dual_conf}.\\
Despite this complexity, some analytical results are available for
the dual user case in \cite{Dush_mmse_dual_conf} for macrodiversity
MMSE and ZF receivers. In \cite{Dush_mmse_dual_conf}, they consider
the statistical properties of the output signal to interference plus
noise ratio (SINR)/signal to noise ratio (SNR) of MMSE and ZF
receivers respectively and obtain high SNR approximations of the
symbol error rate (SER). In \cite{Dush_mrc_conf}, the SER
performance of macrodiversity maximal ratio combining (MRC) has been
exactly derived for arbitrary numbers of users and antenna
configurations. The ergodic sum capacity of the macrodiversity MIMO
multiple access channel is considered in \cite{Dush_sum_cap_jnl}
where tight approximations of ergodic sum capacity are derived in a
compact form. Rayleigh fading is assumed for finite system sizes in
\cite{Dush_mmse_dual_conf, Dush_mrc_conf} and
\cite{Dush_sum_cap_jnl}. One of the analytical techniques used in
\cite{Dush_sum_cap_jnl} is also used here. In
\cite{Dush_sum_cap_jnl}, sum capacity in logarithmic form is
expressed as an exponential and ergodic sum capacity is then written
as the mean of a ratio of quadratic forms. In this work, we have a
very different starting point and consider the characteristic
function (CF) of the SNR/SINR. The exponential in the CF also leads
to a mean of a ratio of quadratic forms. Hence, the two studies
produce similar ratios at this point in the analysis and the same
technique, namely a Laplace type approximation \cite{Lib94}, is
employed both here and in \cite{Dush_sum_cap_jnl} to simplify the
result. Note that the analysis leading up to the ratio of quadratic
forms and following the Laplace approximation is quite specific to
the individual problems considered. As a result,
\cite{Dush_sum_cap_jnl} gives approximate results for ergodic
capacity and here, approximate SER results and SNR/SINR
distributions are obtained. Note that some of the quadratic forms
encountered here are of a different form to those in
\cite{Dush_sum_cap_jnl}. \\
On an another front, an asymptotic large random matrix approach is
employed to derive a deterministic equivalent to the ergodic sum
capacity in \cite{HaLoNa07}. Similarly, an asymptotic approach is
used to study cellular systems with multiple correlated base station
(BS) and user antennas in
\cite{ChImHo09, WeWoCh07}.\\
In this paper, we extend the results in \cite{Dush_mmse_dual_conf}
to more general user and antenna configurations. In particular, the
contributions made are as follows:
%
%
%
\begin{itemize}
  \item [1.] We derive the approximate
  probability distribution function (PDF) and cumulative distribution
  function (CDF) of the output SINR/SNR of MMSE/ZF receivers. The approximate
  cumulative distribution functions are shown to have a remarkably simple
  form as a generalized mixture of exponentials.
  \item [2.] High SNR approximations for the
  SER of MMSE/ZF receivers are derived for a range of modulations
  and these results are used to derive diversity order and array gain
  results. The high SNR results are simple, have a compact form and
  can be used to gain further insights into the effects of channel distribution
  information (CDI) on the performance of macrodiversity MIMO
  systems.
\end{itemize}
%
%
%
%
%
%
The rest of the paper is laid out as follows. Motivational example
for this work is given in Sec. II. Sec. III describes the system
model and receiver types. Sec. IV provides preliminary results which
will be used throughout the paper. The main analysis is given in
Secs. V and VI. Secs. VII and VIII give numerical results and
conclusions, respectively.
\section{Motivational Example}\label{sec:general:motivation}
Consider a hypothetical MIMO spatial multiplexing communication
system with three transmit and three receive antennas. The complex
channel gain, $h_{ik}$, represents the fading coefficient between
transmit antenna $k$ and receive antenna $i$. The average link gain
between the same antennas is given by $P_{ik}$. In the following, we
consider two cases of this MIMO system.
\subsubsection{Macrodiversity System}\label{subsec:general:macro} In
this system, we assume all $P_{ik}$s are different. This case arises
in MIMO systems where both transmit antennas and receive antennas
are widely separated. Then, we consider a particular realization of
the power matrix $\bm{P}=\left(P_{ik}\right)$ for $i,k=1,\dots,3$,
given as
\begin{align}
\bm{P}_{M} &=  \left(
\begin{array}{cccc}
0.3500 & 0.0117  & 0.1225   \\
0.6292 & 0.9282  & 0.0741  \\
0.0208 & 0.0601  & 0.8035   \\
\end{array}  \right).
\label{eq:general:P_M}
\end{align}
\subsubsection{Point-to-Point
System}\label{subsec:general:macro} 
\begin{figure}[!h]
\centering
\includegraphics[scale=0.60]{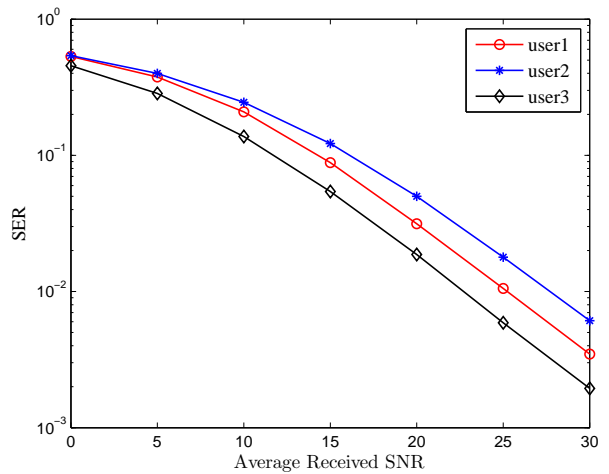}
\caption{Simulated SER results for the macrodiversity case with QPSK
modulation and a ZF receiver. Results are given for all 3 users. }
\label{fig:general:motivation_fig}
\end{figure}

We assume that all the $P_{ik}$
are the same in this system. This case arises in single user MIMO
links with closely spaced transmit and receive antennas. The
point-to-point power matrix is given as
\begin{align}
\bm{P}_{P} &=  \left(
\begin{array}{cccc}
0.3333 & 0.3333  & 0.3333   \\
0.3333 & 0.3333  & 0.3333  \\
0.3333 & 0.3333  & 0.3333   \\
\end{array}  \right).
\label{eq:general:P_P}
\end{align}
Note that the actual values of  $\bm{P}_P$ and $\bm{P}_M$ are
unimportant, it is the comparison between equal and unequal powers
that is relevant. Furthermore, we assume ZF receive combining is
used to spatially separate each independent stream. Each column of
the channel power matrix is normalized so the sum of its power
elements is equal to unity and equal power loading is assumed for
each data stream. This ensures that the average receive power of
each stream is constant and the performance of data streams can be
compared. If we consider the uncoded SER of each stream in both
cases, it is clear that the SER of all three streams is the same in
the point-to-point case due to the symmetry of the channel power
matrix, P. However, in the  macrodiversity case, the picture is not
so clear as shown in Fig. \ref{fig:general:motivation_fig}. Due to
the structure of the power matrix, $\bm{P}_ M$, the curves exhibit
substantial performance gaps in terms of array gain. This effect can
be called the \emph{macrodiversity effect} and is a function of the
slow fading information or the statistical information of the
channel matrix

In this paper, we develop analytical tools to understand the
macrodiversity effect on SER. We also construct some accompanying
statistical measures which may be widely useful for analyzing
macrodiversity systems in the context of the multiuser MIMO multiple
access channel.
\section{System Model}\label{sec:general:system_model}
\begin{figure}[!h]
\centering
\includegraphics[scale=0.85]{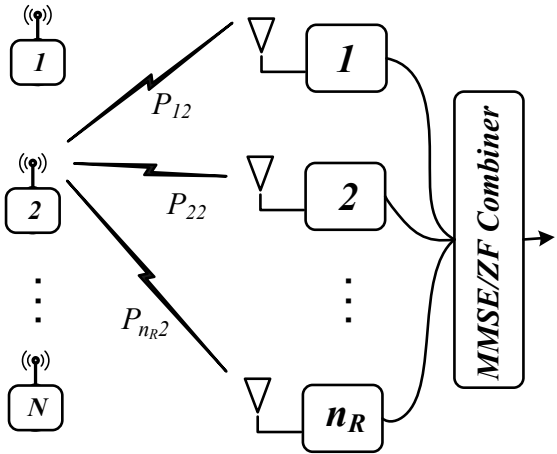}
\noindent \caption{System diagram. To reduce clutter, only paths
from a single source are shown.} \label{fig:general:MU-MIMO_MAC}
\end{figure}

In this section, we present the generic system model which is
considered throughout this paper. The multiuser MIMO system
investigated in this paper consists of $N$ distributed single
antenna users communicating with $n_R$ distributed receive antennas
in an independent flat Rayleigh fading environment. The
$\mathcal{C}^{n_R \times 1}$ receive vector is given by
\vspace{0mm}
\begin{eqnarray}
\bm{r} &=& \bm{Hs} + \bm{n},
\end{eqnarray}
where the $\mathcal{C}^{N \times 1}$ data vector, $\bm{s}=\left(s_1,
s_2, \dots, s_N \right)^T$, contains the transmitted symbols from
the $N$ users and it is normalized, so that $E \left \{ |s_i|^2
\right \}=1$ for $i=1,2,\dots,N$. $\bm{n}$ is the $\mathcal{C}^{n_R
\times 1}$ additive-white-Gaussian-noise (AWGN) vector, $\bm{n} \sim
\mathcal{CN}\left ( \bm{0}, \sigma^2 \bm{I} \right )$, which has
independent entries with  $E \left \{ |n_i|^2 \right \}= \sigma^2$,
for $i=1,2, \dots, n_R$. The channel matrix contains independent
elements, $H_{ik} \sim \mathcal{CN}\left ( 0 , P_{ik} \right )$,
where $E \left \{ |H_{ik}|^2 \right \}= P_{ik}$.
%
%
%
%
A typical macrodiversity MU-MIMO multiple access channel (MAC) is
shown in Fig. \ref{fig:general:MU-MIMO_MAC}, where it is clear that
the geographical spread of users and antennas creates a channel
matrix $\bm{H}$, which has independent entries with different
$P_{ik}$ values. We define the $\mathcal{C}^{n_R \times N}$ matrix,
$\bm{P}=\left \{ P_{ik} \right \}$, which holds the average link
powers due to shadowing, path fading, etc.\\
By assuming that perfect channel state information is available at
the receiver side, we consider a system where channel adaptive
linear combining is performed at the receiver to suppress multiple
access interference \cite{Verdu98}. Therefore, the $\mathcal{C}^{N
\times 1}$ combiner output vector is $\tilde{\bm{r}} = \bm{V}^H
\bm{r}$, where $\bm{V}$ is an $\mathcal{C}^{n_R \times N}$ weight
matrix. In this work, we consider two well-known linear combining
schemes: MMSE and ZF. The structure of $\bm{V}$ and the resulting
output SINR/SNR for MMSE/ZF schemes are well-known and are given
below. Without loss of generality, we assume that the index of the
desired user is $i=1$. The combining vector and output SINR of the
MMSE receiver for user 1 are given by \cite{Gao98, Clark94} as
\vspace*{0mm}
\begin{equation}
\bm{v}_1 = \left(\bm{H}\bm{H}^H + \sigma^2 \bm{I} \right)^{-1}
\bm{h}_1,
\end{equation}
\vspace{2mm}
\begin{equation}
\mbox{SINR} = \bm{h}_1^H \bm{R}^{-1} \bm{h}_1,
\label{eq:general:mmse:sinr}
\end{equation}
where
\begin{equation}
\bm{R} = \sum_{k \neq i}^{N} \bm{h}_k \bm{h}_k^H + \sigma^2 \bm{I},
\end{equation}
and $\bm{H}= \left( \bm{h}_1, \bm{h}_2, \dots, \bm{h}_N \right)$.
Defining $\bm{v}_2, \dots, \bm{v}_N$ similarly gives $\bm{V}= \left(
\bm{v}_1, \bm{v}_2, \dots, \bm{v}_N \right)$. The vectors,
$\bm{h}_k$, clearly play an important role in MMSE combining and it
is useful to define the covariance matrix of $\bm{h}_k$ by $\bm{P}_k
= E \left \{ \bm{h}_k \bm{h}_k^H \right \} = \mbox{diag}(P_{1k},
P_{2k}, \dots , P_{n_Rk})$. From \cite{Winters94, Gore02}, the
combining matrix, $\bm{V}$, and output SNR of the ZF receiver for
$n_R \geq N$ are given by
\begin{equation}
\bm{V} = \bm{H} \left ( \bm{H}^H \bm{H} \right )^{-1}
\end{equation}
and
\begin{equation}
\mbox{SNR} = \frac{1}{\sigma^2 \left[ \left ( \bm{H}^H \bm{H} \right
)^{-1} \right]_{11}}. \label{eq:general:zf:snr}
\end{equation}
\noindent where $\left[\bm{B}\right]_{11}$ indicates the $\left(1,1
\right)^{th}$ element of matrix $\bm{B}$.
\section{Preliminaries}\label{sec:general:preliminaries}
In this section, we state some useful results which will be used
extensively throughout the paper.\\
Let $\bm{A}=\left( a_{ik} \right)$ be an $m \times n$ rectangular
matrix over the commutative ring, $m \leq n $. The permanent of
$\bm{A}$, written $\Perm \left( \bm{A} \right)$, is defined by
\begin{equation}
\Perm \left( \bm{A} \right) = \sum_\sigma a_{1,\sigma_1}
a_{2,\sigma_2} \dots a_{m,\sigma_m},
\end{equation}
\noindent where the summation extends over all one-to-one functions
from $\{1,\dots,m\}$ to $\{1, \dots, n\}$. The sequence
$(a_{1,\sigma_1} a_{2,\sigma_2} \dots a_{m,\sigma_m})$ is called a
diagonal of $\bm{A}$, and the product $a_{1,\sigma_1} a_{2,\sigma_2}
\dots a_{m,\sigma_m}$ is a diagonal product of $\bm{A}$. Thus, the
permanent of $\bm{A}$ is the sum of all diagonal products of
$\bm{A}$.
\begin{Lemma} \cite{Dush_sum_cap_jnl}
Let $\bm{X}$ be an $m\times n$ random matrix with,
\begin{align}
\bm{A} \! &=\! E \left\{ \bm{X} \circ \bm{X}\right\} \triangleq  \!
\left( \!
\begin{array}{ccc} \vspace{1mm}
E \left\{ |X_{11}|^2 \right\} & \ldots  & E \left\{ |X_{1n}|^2 \right\} \\
\vspace{1mm}
E \left\{ |X_{21}|^2 \right\}  & \ldots & E \left\{ |X_{2n}|^2 \right\}\\
\vspace{1mm}
\ldots & \ldots & \ldots  \\
E \left\{ |X_{m1}|^2 \right\}  & \ldots & E \left\{ |X_{mn}|^2 \right\} \\
\end{array} \! \right)\!,
\label{eq:general:expected_matrix_definition}
\end{align}
\noindent where $\circ$ represents the Hadamard product. With this
notation, the following identity holds.
\begin{align}
E \left\{ \left|\bm{X}^H\bm{X}\right|\right\} &= \begin{cases}
\perm \left( \bm{A} \right) & m=n \nonumber \\
\Perm \left( \bm{A} \right) & m>n,
\end{cases}
\end{align}
where $\perm(.)$ and $\Perm \left(.\right)$ are the permanent of a
square matrix and rectangular matrix respectively as defined in
\cite{Minc78}. \label{lemma:general:expected_square_determinant}
\end{Lemma}
%
%
%
%
%
\begin{Corollary}
\cite{Dush_sum_cap_jnl} Let $\bm{X}$ be an $m\times n$ random matrix
with, $E \left\{ \bm{X} \circ \bm{X}\right\} = \bm{A}$, where
$\bm{A}$ is an $m\times n$ deterministic matrix and $m>n$. If the
$m\times m$ deterministic matrix $\bm{\Sigma}$ is diagonal, then the
following identity holds,
\begin{align}
\begin{split}
E \left\{ \left|\bm{X}^H\bm{\Sigma}\bm{X}\right| \right\} = \Perm
\left( \bm{\Sigma} \bm{A} \right).
\end{split}
\end{align}
\label{corollary:general:expected_diagonal_rectangular_determinant}
\end{Corollary}
%
%
%
%
Next, we present three axiomatic identities for permanents
\cite{Dush_sum_cap_jnl}.
\begin{itemize}
\item \emph{Axiom 1}: For an empty matrix, $\bm{A}$,
\begin{align}
\Perm \left(\bm{A} \right)&=1. \label{eq:general:axiom3}
\end{align}
\item \emph{Axiom 2}: Let $\bm{A}$ be an arbitrary $m\times n$ matrix, then
\begin{align}
\sum_{\sigma} \Perm \left( \left(\bm{A} \right)^{\sigma_{0,n}}
\right) &= \sum_{\sigma} \Perm \left( \left(\bm{A}
\right)_{\sigma_{0,m}} \right) = 1.
\label{eq:general:axiom1}
\end{align}
\item \emph{Axiom 3}: Let $\bm{A}$ be an arbitrary $m\times n$ matrix, then
\begin{align}
\sum_{\sigma} \Perm \left( \left( \bm{A}\right)_{\sigma_{k,m}}
\right) &= \sum_\sigma \Perm \left( \left(
\bm{A}\right)^{\sigma_{k,n}} \right), \label{eq:general:axiom2}
\end{align}
\end{itemize}
\noindent where $\sigma_{k,n}$ is an ordered subset of $\left \{
n\right\}=\left\{1, \dots, n \right\}$ of length $k$ and the
summation over all such subsets.
$\left.\bm{X}\right._{\sigma_{\ell,n}}$ denotes the principal
submatrix of $\bm{X}$ formed by taking only the rows and
columns indexed by $\sigma_{\ell,n}$.\\
In general, $\left.\bm{X}\right._{\sigma_{\ell,n}}^{\mu_{\ell,n}}$
denotes the submatrix of $\bm{X}$ formed by taking only the rows and
columns indexed by $\sigma_{\ell,n}$ and $\mu_{\ell,n}$
respectively, where $\sigma_{\ell,n}$ and $\mu_{\ell,n}$ are length
$\ell$ subsets of $\{1, 2, \dots ,n\}$. If either $\sigma_{\ell,n}$
or $\mu_{\ell,n}$ contain the complete set, the corresponding
subscript/superscript may be dropped. When
$\sigma_{\ell,n}=\mu_{\ell,n}$, only one
subscript/superscript may be shown for brevity.\\
\section{ZF Analysis}\label{sec:general:zf}
In this section, we derive an approximate CDF for the output SNR of
a ZF receiver, a high SNR approximation to SER and also consider
some special cases. The following PDFs for the columns of the
channel matrix are used throughout the analysis.
\begin{equation}
f(\bm{h}_k) = \frac{1}{\pi^{n_R} \left|\bm{P}_k\right|}
e^{-\bm{h}_k^H \bm{P}_k^{-1} \bm{h}_k },
 \label{eq:general:PDFs}
\end{equation}
\noindent for $k=1,2, \dots,N$.
\subsection{CDF Approximations}\label{subsec:general:zf:CDF_approximations}
The output SNR of a ZF receiver in \dref{eq:general:zf:snr} can be
written as
\begin{align}
\tilde{Z} &= \frac{1}{\sigma^2} \bm{h}_1^H \left(\bm{I} - \bm{H}_2
\left(\bm{H}_2^H \bm{H}_2 \right)^{-1}\bm{H}_2^H \right)\bm{h}_1 \\
&= \frac{1}{\sigma^2} \bm{h}_1^H \bm{M} \bm{h}_1,
\label{eq:general:zf:snr1}
\end{align}
where $\bm{M}=\bm{I} - \bm{H}_2 \left(\bm{H}_2^H \bm{H}_2
\right)^{-1}\bm{H}_2^H$ and $\bm{H}_2$ is $\bm{H}$, with $\bm{h}_1$
removed. Following the analysis in \cite{Dush_mmse_dual_conf}, the
characteristic function (CF) of $\tilde{Z}$ is given by
\begin{eqnarray}
\phi_{\tilde{Z}}(t) &=& E \left \{ e^{jt\tilde{Z}} \right \} = E
\left \{ e^{\frac{jt}{\sigma^2} \bm{h}_1^H \bm{M} \bm{h}_1 }
\right\}. \label{eq:general:zf:CF}
\end{eqnarray}
\noindent Conditioning on $\bm{H}_2$, the expectation over
$\bm{h}_1$ in \dref{eq:general:zf:CF} can be solved as in
\cite{Dush_mmse_dual_conf} to obtain
\begin{equation}
\phi_{\tilde{Z}}(t | \bm{H}_2) = \frac{1}{\left| \bm{I} - jt
\frac{1}{\sigma^2}\bm{M} \bm{P}_1 \right|}.
\label{eq:general:zf:CF1}
\end{equation}
Substituting for $\bm{M}$ in \dref{eq:general:zf:CF1} gives
\begin{align}
\phi_{\tilde{Z}}(t | \bm{H}_2) &= \frac{1}{\left| \bm{I} -
\frac{jt}{\sigma^2}\bm{P}_1 +  \frac{jt}{\sigma^2} \bm{H}_2
\left(\bm{H}_2^H \bm{H}_2 \right)^{-1}\!\! \bm{H}_2^H \bm{P}_1\right|}\\
&=\frac{1}{\left|\bm{D}\right|\left| \bm{I}  + \frac{jt}{\sigma^2}
\bm{H}_2 \left(\bm{H}_2^H \bm{H}_2 \right)^{-1}\!\!\bm{H}_2^H
\bm{P}_1
\bm{D}^{-1}\right|} \\
&=\frac{\left|\bm{H}_2^H \bm{H}_2\right|}{\left|\bm{D}\right|\left|
\bm{H}_2^H \bm{H}_2  + \frac{jt}{\sigma^2} \bm{H}_2^H \bm{P}_1
\bm{D}^{-1}\bm{H}_2\right|}, \label{eq:general:zf:CF2}
\end{align}
\noindent where $\bm{D}= \bm{I} - \frac{1}{\sigma^2}jt\bm{P}_1$.
Simplifying \dref{eq:general:zf:CF2} using the result $\bm{I} +
\frac{jt}{\sigma^2}\bm{P}_1 \bm{D}^{-1} =\bm{D}^{-1}$ gives
\begin{align}
\phi_{\tilde{Z}}(t | \bm{H}_2) &= \frac{\left|\bm{H}_2^H
\bm{H}_2\right|}{\left|\bm{D}\right|\left|\bm{H}_2^H \bm{D}^{-1}
\bm{H}_2\right|}. \label{eq:general:zf:CF3}
\end{align}
\noindent The full CF can then be obtained by averaging the
conditional CF in \dref{eq:general:zf:CF3}, to give
\begin{align}
\phi_{\tilde{Z}}(t) &= \frac{1}{\left|\bm{D}\right|} E\left \{
\frac{\left|\bm{H}_2^H \bm{H}_2\right|}{\left|\bm{H}_2^H \bm{D}^{-1}
\bm{H}_2\right|} \right \}, \label{eq:general:zf:CF4}
\end{align}
where expectation is over $\bm{H}_2$. An exact analysis of
\dref{eq:general:zf:CF4} is extremely cumbersome. However, for the
dual source scenario where $N=2$, \dref{eq:general:zf:CF4} can be
solved in closed form \cite{Dush_mmse_dual_jnl}. Even for $N=2$, the
resulting exact expressions are complex. Hence, for arbitrary $N$ we
use a Laplace type approximation as in \cite{Dush_mmse_dual_conf} to
approximate and simplify the CF. This approximation has some
motivation in the work of \cite{Lib94} and \cite{Dush_sum_cap_jnl,
mythesis}. It can also be thought of as a first order delta
expansion \cite{StOr94}. Further insight into the use of the Laplace
approximation can be gained from the case where $n_R$ is large and
N=2. Here, $\bm{H}_2$ is a column vector and the numerator and
denominator in \dref{eq:general:zf:CF5} are standard quadratic
forms. Normalizing both numerator and denominator by dividing by
$n_R$ leads to a ratio where both quadratic forms tend to constants
as long as the conditions of a version of the weak law of large
numbers hold. In this case, \dref{eq:general:zf:CF5} becomes
asymptotically exact. Hence, the stabilizing effect of averaging in
the numerator and denominator is the motivation for the use of the
Laplace approximation. This approach gives
\begin{align}
\phi_{\tilde{Z}}(t) &\simeq  \frac{1}{\left|\bm{D}\right|}
\frac{E\left \{ \left|\bm{H}_2^H \bm{H}_2\right| \right\}}{E\left \{
\left|\bm{H}_2^H \bm{D}^{-1} \bm{H}_2\right| \right \} }.
\label{eq:general:zf:CF5}
\end{align}
It is worth noting here that if the diagonal matrix, $\bm{P}_1$, is
a scaled identity matrix, the approximation in
\dref{eq:general:zf:CF5} becomes exact regardless of the power
matrix of $\bm{H}_2$. In contrast to microdiversity ZF, when
$\bm{P}_1$ is not a scaled identity matrix, then $\bm{D}$ in
\dref{eq:general:zf:CF4} is not a scaled identity and cannot be
factorized out of the determinant. As a result, the CF in
\dref{eq:general:zf:CF4} depends on the expected value over
$\bm{H}_2$ and so the statistical performance changes according to
the power matrix of $\bm{H}_2$ (i.e., according to the
macrodiversity power profile of the interference). Applying Lemma
\ref{lemma:general:expected_square_determinant} and Corollary
\ref{corollary:general:expected_diagonal_rectangular_determinant} to
\dref{eq:general:zf:CF5} gives
\begin{align}
\phi_{\tilde{Z}}(t) &\simeq  \frac{\Perm \left(\bm{Q}_2
\right)}{\left|\bm{D}\right|\Perm \left(\bm{D}^{-1} \bm{Q}_2
\right)}, \label{eq:general:zf:CF6}
\end{align}
where the $n_R \times \left(N-1\right)$ matrix $\bm{Q}_2$ is defined
by $\bm{Q}_2 = E \left\{ \bm{H}_2 \circ \bm{H}_2\right\}$.
$\bm{P}=\left( \bm{p}_1 \bm{Q}_2 \right)$, and
$\bm{p}_1=\left(P_{11}, P_{21}, \dots, P_{n_R 1}\right)^T$. From
Appendix \ref{app:general:zf_denom_expect}, the denominator in
\dref{eq:general:zf:CF6} can be expanded as
\begin{align}
\left|\bm{D}\right| \Perm \left(\bm{D}^{-1} \bm{Q}_2 \right) =
\sum_{i=0}^{L} \left(-jt\right)^i \tilde{\varphi}_i,
\label{eq:general:zf:denominator:expectation2}
\end{align}
\noindent where
\begin{align}
\tilde{\varphi}_i &= \sum_{\sigma} \Tr_i \left(
\left(\bm{P}_1\right)_{\bar{\sigma}_{L,n_R}} \right) \perm \left(
\left( \bm{Q}_2 \right)_{\sigma_{N-1,n_R}}^{\{N-1\}} \right)
\left(\sigma^2 \right)^{-i},
\label{eq:general:zf:denominator:expectation:varphi}
\end{align}
and $\Tr_i\left(.\right)$ are elementary symmetric functions defined
in \cite[1.2.12]{Horn85}. Since $\perm \left( \left( \bm{Q}_2
\right)_{\sigma_{N-1,n_R}}^{\{N-1\}}\right)$ is independent of $t$,
it is clear from \dref{eq:general:zf:denominator:expectation:varphi}
that $\left|\bm{D}\right| \Perm \left(\bm{D}^{-1} \bm{Q}_2 \right)$
is a polynomial in $t$ of degree $L$. Hence,
\dref{eq:general:zf:CF6} becomes
\begin{align}
\phi_{\tilde{Z}}(t) &\simeq  \frac{\Perm \left(\bm{Q}_2
\right)}{\sum_{i=0}^L  \left(-jt\right)^i \tilde{\varphi}_i}  \\
&=  \frac{\Perm \left(\bm{Q}_2 \right)}{\tilde{\varphi}_L
\sum_{i=0}^L
\left( \frac{\tilde{\varphi}_i}{\tilde{\varphi}_L} \right) \left(-jt\right)^i} \label{eq:general:zf:CF101} \\
&= \frac{\Perm \left(\bm{Q}_2 \right)}{\tilde{\varphi}_L
\prod_{i=1}^L
\left( \tilde{\omega}_i - jt \right)} \\
&=  \frac{\Perm \left(\bm{Q}_2 \right)}{\tilde{\varphi}_L}
\sum_{i=1}^L \frac{\tilde{\eta}_i}{\left. \tilde{\omega}_i - jt
\right.}, \label{eq:general:zf:CF11}
\end{align}
\noindent where $\tilde{\omega}_i$ are the roots of the denominator
polynomial in \dref{eq:general:zf:CF101}. These roots can be
computed using standard root finding programs. Note that the roots
are all positive, $\tilde{\omega}_i
> 0$ for all $i$, from Descarte's rule of signs and
\begin{align}
\tilde{\eta}_i &= \frac{1}{\prod_{k \neq i}^{n_R}
\left(\tilde{\omega}_k - \tilde{\omega}_i \right)}.
\label{eq:general:zf:etai0}
\end{align}
It is clear from \dref{eq:general:zf:CF6} that
$\phi_{\tilde{Z}}(0)=1$, since $\bm{D}=\bm{I}$ when $t=0$.
Therefore, the CF will produce a valid PDF after inversion
\cite{Cramer45}. From \cite{Proakis01}, the PDF and CDF of
$\tilde{Z}$ are given by
\begin{equation}
f_{\tilde{Z}}(z) = \frac{1}{2\pi}\int_{-\infty}^\infty
\phi_{\tilde{Z}}(t) e^{-jtz} dt, \label{eq:general:zf:PDF}
\end{equation}
\begin{equation}
F_{\tilde{Z}}(z) = \frac{1}{2\pi} \int_0^z \int_{-\infty}^\infty
\phi_{\tilde{Z}}(t) e^{-jtx} dt dx. \label{eq:general:zf:CDF}
\end{equation}
Substituting \dref{eq:general:zf:CF11} in \dref{eq:general:zf:PDF}
gives the approximate PDF
\begin{align}
\hat{f}_{\tilde{Z}}(z) &= \frac{\Perm \left(\bm{Q}_2 \right)}{2\pi
\tilde{\varphi}_L} \sum_{i=1}^L \tilde{\eta}_i
\int_{-\infty}^{\infty} \frac{e^{-jtz}}{\left. \tilde{\omega}_i - jt
\right.} dt. \label{eq:general:zf:PDF1}
\end{align}
Applying the integral identity from \cite[eq. 7, 3.382]{GradRzy00},
%
%
we obtain the approximate PDF of $\tilde{Z}$ as
\begin{align}
\hat{f}_{\tilde{Z}}(z) &=  \frac{\Perm \left(\bm{Q}_2 \right)}{
\tilde{\varphi}_L} \sum_{i=1}^L \tilde{\eta}_i
e^{-\tilde{\omega}_iz}, \label{eq:general:zf:PDF2}
\end{align}
and the approximate CDF of $\tilde{Z}$ becomes
\begin{align}
\hat{F}_{\tilde{Z}}(z) &=  \frac{\Perm \left(\bm{Q}_2 \right)}{
\tilde{\varphi}_L} \sum_{i=1}^L
\frac{\tilde{\eta}_i}{\tilde{\omega}_i} \left( 1 -
e^{-\tilde{\omega}_iz} \right). \label{eq:general:zf:CDF2}
\end{align}
The final PDF approximation in \dref{eq:general:zf:PDF2} has a
remarkably simple form as a generalized mixture of $L$ exponentials
where $L=n_R-N+1$. In section VII, a numerical example is given to
show that the performance of a macrodiversity system is very hard to
predict without the relevant analytical performance metrics. In the
case of (38), we are able to approximate outage probabilities and
this is a metric of particular interest in cell-edge scenarios where
outage is of particular concern.
\subsection{Special Cases}\label{subsec:general:zf:special_cases} In
this section we present the special case where $n_R=N$, i.e., the
system schedules as many simultaneous users as the number of receive
antennas. In this particular scenario, the ZF CDF analysis in Sec.
\ref{subsec:general:zf:CDF_approximations} has an intriguing form.
From \dref{eq:general:zf:CF6}, the CF of $\tilde{Z}$ becomes
\begin{align}
\phi_{\tilde{Z}}(t) &\simeq \frac{\Perm \left(\bm{Q}_2
\right)}{\left|\bm{D}\right|\Perm \left(\bm{D}^{-1} \bm{Q}_2
\right)}. \label{eq:general:zf:special_case1}
\end{align}
When $n_R=N$, the denominator of \dref{eq:general:zf:special_case1}
simplifies to give,
\begin{align}\label{eq:general:zf:special_case2}
\left|\bm{D}\right|\Perm \left(\bm{D}^{-1} \bm{Q}_2 \right) &=
\sum_{i=1}^{n_R} \left( 1 - \frac{jt}{\sigma^2} P_{i1} \right) \perm
\left( \bm{Q}_{i2} \right),
\end{align}
\noindent where $\bm{Q}_{i2}$ is $\bm{Q}_{2}$ with the
$i\textsuperscript{th}$ row removed. Then,
\dref{eq:general:zf:special_case1} simplifies to
\begin{align}
\phi_{\tilde{Z}}(t) &\simeq \frac{\Perm \left(\bm{Q}_2
\right)}{\sum_{i=1}^{n_R} \perm \left( \bm{Q}_{i2} \right) -
\frac{jt}{\sigma^2} \sum_{i=1}^{n_R} P_{i1} \perm \left( \bm{Q}_{i2}
\right)}, \\
&=\frac{\Perm \left(\bm{Q}_2 \right)}{\Perm \left(\bm{Q}_2 \right) -
\frac{jt}{\sigma^2} \perm \left( \bm{P} \right) }.
\label{eq:general:zf:special_case3}
\end{align}
Inverting the CF expression in \dref{eq:general:zf:special_case3}
gives the approximate PDF of $\tilde{Z}$ as the simple exponential
\begin{align}\label{eq:general:zf:special_case4}
\hat{f}_{\tilde{Z}}(z) &= \sigma^2 \theta e^{-\sigma^2 \theta z},
\end{align}
\noindent where $\theta=\Perm \left(\bm{Q}_2 \right)/\perm
\left(\bm{P} \right)$.
\subsection{High SNR Approximations}\label{subsec:general:zf:high_snr_approximations}
The CF in \dref{eq:general:zf:CF4} is a ratio of determinants, where
$\bm{D}= \bm{I} - \frac{1}{\sigma^2}jt\bm{P}_1$. As the SNR grows,
$\sigma^2\rightarrow 0$ and keeping only the dominant power of
$\sigma^2$ in \dref{eq:general:zf:CF4} gives
\begin{align}
\phi_{\tilde{Z}}(t) &= \tilde{K}_0 \left( \frac{\sigma^2}{-jt}
\right)^{n_R-N+1}, \label{eq:general:zf:high_snr:CF1}
\end{align}
\noindent where
\begin{align}
\tilde{K}_0 &= \frac{1}{\left|\bm{P}_1\right|} E\left \{
\frac{\left|\bm{H}_2^H \bm{H}_2\right|}{\left|\bm{H}_2^H
\bm{P}_1^{-1} \bm{H}_2\right|} \right \}.
\label{eq:general:zf:high_snr:K0}
\end{align}
Following the MGF based approach in \cite{SiAl00}, the SER of a
macrodiversity ZF receiver can be evaluated for $M$-PSK modulation
as
\begin{align}
\tilde{P}_s = \frac{1}{\pi} \int_0^T \mathcal{M}_{\tilde{Z}} \left(
-\frac{g}{\sin^2 \theta} \right )
d\theta.\label{eq:general:zf:high_snr:ser1}
\end{align}
\noindent where $\mathcal{M}_{\tilde{Z}} \left(s\right) =
\phi_{\tilde{Z}} \left( -js \right)$, $g=\sin^{2}\left(\pi / M
\right)$ and $T=\frac{\left(M-1\right)\pi}{M}$. Note that linear
combinations of equations of the form given in
\dref{eq:general:zf:high_snr:ser1} also give SERs for $M$-QAM in the
usual way \cite{SiAl00}. Substituting
\dref{eq:general:zf:high_snr:CF1} in
\dref{eq:general:zf:high_snr:ser1} gives
\begin{align}
\tilde{P}^{\infty}_s = \left(\tilde{G}_a
\bar{\gamma}\right)^{-\tilde{G}_d} +
o\left(\bar{\gamma}^{-\tilde{G}_d} \right),
\label{eq:general:zf:high_snr:ser2}
\end{align}
\noindent where $o\left(.\right)$ is the standard little-o notation
and the average SNR is $\bar{\gamma}=\frac{1}{\sigma^{2}}$. The
diversity gain and array gain in \dref{eq:general:zf:high_snr:ser2}
are given by
\begin{align}
\tilde{G}_d&=n_R-N+1, \hspace{10mm} \tilde{G}_a =\left( \tilde{K}_0
\tilde{\mathcal{I}} \right)^{-1/(n_R-N+1)}, \nonumber
\end{align}
where $\tilde{\mathcal{I}}$ is given by
\begin{align}
\tilde{\mathcal{I}} = \frac{1}{\pi} \int_0^T \left( \frac{
\sin^2\theta}{g} \right)^{\left(n_R-N+1\right)} d\theta.
\label{eq:general:zf:high_snr:integral}
\end{align}
The high SNR expression derived in
\dref{eq:general:zf:high_snr:ser2} will be exact, if and only if
$\tilde{K}_0$ is exact. An exact calculation of $\tilde{K}_0$ for
the $N=2$ case is presented in \cite{Dush_mmse_dual_jnl} and shown
to have a complex expression. This work suggests that in the general
case an exact calculation is likely to be either excessively
complicated or intractable. Hence, in this work, we use a Laplace
type approximation for $\tilde{K}_0$ in
\dref{eq:general:zf:high_snr:K0} to obtain a more compact and
insightful expression. Hence, we use the following approximation
\begin{align}
\tilde{K}_0 &\simeq \frac{1}{\left|\bm{P}_1\right|}  \frac{E\left \{
\left|\bm{H}_2^H \bm{H}_2\right| \right \} }{ E \left\{
\left|\bm{H}_2^H \bm{P}_1^{-1} \bm{H}_2\right|
 \right\} }. \label{eq:general:zf:high_snr:K0:1}
\end{align}
Using Lemma \ref{lemma:general:expected_square_determinant},
\dref{eq:general:zf:high_snr:K0:1} is given by
\begin{align}
\tilde{K}_0 &\simeq  \frac{\Perm \left(\bm{Q}_2
\right)}{\left|\bm{P}_1\right|\Perm \left(\bm{P}_1^{-1} \bm{Q}_2
\right)}. \label{eq:general:zf:high_snr:K0:2}
\end{align}
Note that when $N=2$, approximate $\tilde{K}_0$ has simpler
expression \cite{Dush_mmse_dual_conf}, which gives
\begin{align}
\tilde{K}_0 &\simeq  \frac{\Tr \left(\bm{P}_2
\right)}{\left|\bm{P}_1\right| \Tr\left(\bm{P}_1^{-1} \bm{P}_2
\right)}. \label{eq:general:zf:high_snr:K0:dual_user}
\end{align}
The high SNR SER approximation in \dref{eq:general:zf:high_snr:ser2}
has the useful property that all the dependence on P is encapsulated
in the $\tilde{K_0}$ metric in \dref{eq:general:zf:high_snr:K0:2}.
Hence, $\tilde{K_0}$ acts as a stand-alone performance metric as
shown in the numerical example in Sec. VII. This feature has
implications for systems where only long-term CSI is available for
scheduling. Here, $\tilde{K_0}$ can be used as a scheduling metric
\cite{SmBa12} as it is a one-to-one function of the approximate SER.
Such situations include systems with rapidly changing channels,
systems where CSI exchange is too expensive and systems with large
numbers of sources and/or receivers. In all these cases, long term
CSI based scheduling may be preferable due to the overheads, delays
and errors implicit in obtaining instantaneous CSI \cite{BaVi07}.
\section{MMSE Analysis}\label{sec:general:mmse}

\subsection{CDF Approximations}\label{subsec:general:mmse:CDF_approximations}
In this section, we derive the approximate CDF of the output SINR of
an MMSE receiver and a high SNR approximation to the SER. Let Z be
the output SINR of an MMSE receiver given by
\dref{eq:general:mmse:sinr}. Following the same procedure as in the
ZF analysis, the CF of Z is
\begin{eqnarray}
\phi_{Z}(t) &=& E \left \{ e^{jtZ} \right \} = E \left \{ e^{jt
\bm{h}_1^H \bm{R}^{-1} \bm{h}_1 } \right\}.
\label{eq:general:mmse:CF}
\end{eqnarray}
Next, the CF conditioned on $\bm{H}_2$ becomes
\cite{Dush_mmse_dual_conf}
\begin{equation}
\phi_{Z}(t | \bm{H}_2) = \frac{1}{\left| \bm{I} - jt \bm{R}^{-1}
\bm{P}_1 \right|}. \label{eq:general:mmse:CF2}
\end{equation}
Since $\bm{R}=\sigma^2 \bm{I} + \bm{H}_2 \bm{H}_2^H$, the
conditional CF in \dref{eq:general:mmse:CF2} becomes
\begin{align}
\phi_{Z}(t | \bm{H}_2) &= \frac{1}{\left| \bm{I} - jt \left(\sigma^2
\bm{I} +
\bm{H}_2 \bm{H}_2^H\right)^{-1} \bm{P}_1 \right|} \\
&= \frac{\left|\sigma^2 \bm{I} + \bm{H}_2
\bm{H}_2^H\right|}{\left|\sigma^2 \bm{I} -jt \bm{P}_1 + \bm{H}_2
\bm{H}_2^H\right|}. \label{eq:general:mmse:CF3}
\end{align}
Using the determinant identity, $\left|\bm{I} +
\bm{X}\bm{X}^H\right|=\left|\bm{I} + \bm{X}^H\bm{X}\right|$, where
the rank of the identity matrix is obvious from the context, in
\dref{eq:general:mmse:CF3} along with some simple algebra, we get
\begin{align}
\phi_{Z}(t | \bm{H}_2) &= \frac{\left|\sigma^2 \bm{I} + \bm{H}_2^H
\bm{H}_2\right|}{\left|\bm{D}\right|\left|\sigma^2 \bm{I} +
\bm{H}_2^H \bm{D}^{-1}\bm{H}_2\right|}, \label{eq:general:mmse:CF4}
\end{align}
\noindent where $\bm{D}= \bm{I} - \frac{1}{\sigma^2}jt\bm{P}_1$.
Note the similarity of \dref{eq:general:mmse:CF4} with
\cite[eq.~14]{Dush_mmse_dual_conf}. Then, the full CF can be solved
by averaging the conditional CF in \dref{eq:general:mmse:CF4} over
$\bm{H}_2$. Hence,
\begin{align}
\phi_{Z}(t) &= \frac{1}{\left|\bm{D}\right|} E \left\{
\frac{\left|\sigma^2 \bm{I} + \bm{H}_2^H
\bm{H}_2\right|}{\left|\sigma^2 \bm{I} + \bm{H}_2^H
\bm{D}^{-1}\bm{H}_2\right|} \right\}. \label{eq:general:mmse:CF5}
\end{align}
Using a similar approach as in the ZF analysis we approximate
\dref{eq:general:mmse:CF5} to get
\begin{align}
\phi_{Z}(t) &\simeq \frac{1}{\left|\bm{D}\right|}  \frac{E \left\{
\left|\sigma^2 \bm{I} + \bm{H}_2^H \bm{H}_2\right| \right\} }{E
\left\{ \left|\sigma^2 \bm{I} + \bm{H}_2^H
\bm{D}^{-1}\bm{H}_2\right| \right\}  }. \label{eq:general:mmse:CF6}
\end{align}
In Appendix \ref{app:general:mmse_expectation1} we obtain the
expectation in the numerator of \dref{eq:general:mmse:CF6} as
\begin{align}
E \left\{ \left|\sigma^2 \bm{I} + \bm{H}_2^H \bm{H}_2\right| \right
\} \!\! &=\!\! \sum_{k=0}^{N-1} \!\!\sum_{\sigma} \Perm \left(
\left(\bm{Q}_2 \right)^{\sigma_{k,N-1}} \right) \left(\sigma^2
\right)^{N-k-1}. \label{eq:general:mmse_numerator:expectation}
\end{align}
\noindent Appendix \ref{app:general:mmse_expectation2_app} gives the
denominator of \dref{eq:general:mmse:CF6} as
\begin{align}
\left|\bm{D}\right| E \left\{ \left|\sigma^2 \bm{I} + \bm{H}_2^H
\bm{D}^{-1} \bm{H}_2\right| \right \} &= \sum_{i=0}^{n_R}
\left(-jt\right)^{i} \varphi_i,
\label{eq:general:mmse_denominator:expectation1}
\end{align}
\noindent where
\begin{align}
\varphi_i &=  \sum_{k=0}^{N-1} \hat{\varphi}_{ik} \left(\sigma^2
\right)^{N-i-k-1},
\label{eq:general:mmse_denominator:expectation:varphi1}
\end{align}
\noindent and $\hat{\varphi}_{ik}$ is given in
\dref{eq:general:mmse_denominator:expectation:acute_varphi1}.
Substituting \dref{eq:general:mmse_numerator:expectation} and
\dref{eq:general:mmse_denominator:expectation1} in
\dref{eq:general:mmse:CF6} we get,
\begin{align}
\phi_{Z}(t) &\simeq \frac{\sum_{k=0}^{N-1} \!\!\sum_{\sigma} \Perm
\left( \left(\bm{Q}_2 \right)^{\sigma_{k,N-1}} \right)
\left(\sigma^2 \right)^{N-k-1}}{\sum_{i=0}^{n_R}
\left(-jt\right)^{i} \varphi_i } \\
 &= \frac{\Theta \left(\bm{Q}_2 \right)}{\varphi_{n_R} \sum_{i=0}^{n_R}
\left( \frac{\varphi_i}{\varphi_{n_R}}\right) \left(-jt\right)^{i} }
\\
&= \frac{\Theta \left(\bm{Q}_2 \right)}{ \varphi_{n_R}
\prod_{i=1}^{n_R} \left(\omega_i -jt\right) },
\label{eq:general:mmse:CF7}
\end{align}
\noindent where $\tilde{\omega}_i > 0$ for all $i$ is from
Descarte's rule of signs and
\begin{align}
\Theta \left(\bm{Q}_2 \right) = \sum_{k=0}^{N-1} \!\!\sum_{\sigma}
\Perm \left( \left(\bm{Q}_2 \right)^{\sigma_{k,N-1}} \right)
\left(\sigma^2 \right)^{N-k-1}. \label{eq:general:mmse:theta}
\end{align}
The final expression for $\phi_Z(t)$ then becomes
\begin{align}
\phi_{Z}(t) &\simeq \frac{\Theta \left(\bm{Q}_2
\right)}{\varphi_{n_R}} \sum_{i=1}^{n_R} \frac{\eta_i}{\omega_i
-jt}, \label{eq:general:mmse:CF8}
\end{align}
\noindent where
\begin{align}
\eta_i &= \frac{1}{\prod_{k \neq i}^{n_R} \left(\omega_k - \omega_i
\right)}. \label{eq:general:mmse:etai0}
\end{align}
As in the ZF analysis, the PDF and CDF of $Z$ can be computed
%
%
using the identity in \cite[eq. 7, 3.382]{GradRzy00}. Finally we get
the approximate PDF of $Z$ as
\begin{align}
\hat{f}_{Z}(z) &=  \frac{ \Theta \left(\bm{Q}_2 \right)}{
\varphi_{n_R}} \sum_{i=1}^{n_R} \eta_i e^{-\omega_iz},
\label{eq:general:mmse:PDF2}
\end{align}
and the CDF of $Z$ becomes
\begin{align}
\hat{F}_{Z}(z) &=  \frac{\Theta \left(\bm{Q}_2 \right)}{
\varphi_{n_R}} \sum_{i=1}^{n_R} \frac{\eta_i}{\omega_i} \left( 1 -
e^{-\omega_iz} \right). \label{eq:general:mmse:CDF2}
\end{align}
In contrast to \dref{eq:general:zf:PDF2}, where the ZF SNR is a
generalized mixture of $L$ exponentials, \dref{eq:general:mmse:PDF2}
can be identified as a generalized mixture of $n_R \geq L$
exponentials. Since the MMSE SINR has more mixing parameters ($n_R$
rather than $L$) it might be expected that these increased degrees
of freedom will result in a better approximation. Alternatively, the
more concise ZF result, which provides a lower bound on the MMSE
performance, can be used to provide a simpler expression for use in
system design and understanding.
\subsection{High SNR Approximations}\label{subsec:general:mmse:high_snr_approximations}
The CF in \dref{eq:general:mmse:CF6} is a ratio of determinants. As
the SNR grows, $\sigma^2\rightarrow 0$ and keeping only the dominant
power of $\sigma^2$ in \dref{eq:general:mmse:CF6} gives
\begin{align}
\phi(t) &= K_0\left(-jt\right) \left( \frac{\sigma^2}{-jt}
\right)^{n_R-N+1}, \label{eq:general:mmse:high_snr:CF1}
\end{align}
\noindent where
\begin{align}
K_0\left(s\right) &= \frac{1}{\left|\bm{P}_1\right|} E\left \{
\frac{\left|\bm{H}_2^H \bm{H}_2\right|}{\left| \bm{H}_2^H
\bm{P}_1^{-1} \bm{H}_2 + s\bm{I} \right|} \right \}.
\label{eq:general:mmse:high_snr:K0}
\end{align}
Hence, from \dref{eq:general:zf:high_snr:ser1}, the SER at high SNR
becomes
\begin{align}
\!\!\!P^{\infty}_s &= \frac{1}{\pi} \! \int_0^T \!\!\left(
\frac{\sigma^2 \sin^2 \theta}{g}\right)^{n_R-N+1} \!\!\!K_0
\left(\frac{g}{\sin^2 \theta} \right) d\theta.
\label{eq:general:mmse:high_snr:ser1}
\end{align}
As in the ZF analysis, an exact calculation of $K_0$ appears
difficult and we use the Laplace-type approximation again to give
\begin{align}
K_0\left(s\right) &\simeq \frac{1}{\left|\bm{P}_1\right|}  \frac{E
\left\{ \left|\bm{H}_2^H \bm{H}_2\right| \right\} }{E \left\{
\left|s \bm{I} + \bm{H}_2^H \bm{P}_1^{-1}\bm{H}_2\right| \right\} }.
\label{eq:general:mmse:high_snr:K01}
\end{align}
From Lemma \ref{lemma:general:expected_square_determinant} and
\dref{eq:general:mmse_numerator:expectation}, we have
\begin{align}
K_0\left(s\right) &\simeq \frac{\Perm \left(\bm{Q}_2
\right)}{\left|\bm{P}_1\right| \left( \sum_{i=0}^{N-1} \!\! \zeta_i
\left.s \right.^{N-i-1} \right)}, \\
&=  \frac{\Perm \left(\bm{Q}_2 \right) \left( \prod_{i=1}^{N-1}
\vartheta_i\right) }{\left|\bm{P}_1\right| \left( \prod_{i=1}^{N-1}
\vartheta_i\right)\left. \prod_{i=1}^{N-1} \!\! \left( \vartheta_i
+s
\right) \right.},\\
&=  \frac{\Perm \left(\bm{Q}_2 \right)}{\left|\bm{P}_1\right|
\zeta_{N-1}} \sum_{i=1}^{N-1} \frac{\chi_i}{\vartheta_i +s},
\label{eq:general:mmse:high_snr:K02}
\end{align}
\noindent where $\zeta_i=\sum_{\sigma} \Perm \left(
\left(\bm{P}_1^{-1} \bm{Q}_2 \right)^{\sigma_{i,N-1}} \right)$ and
$-\vartheta_i$ are the roots of $\sum_{i=0}^{N-1} \!\! \zeta_i
\left.s \right.^{N-i-1}$. Since $\zeta_{N-1}=\Perm
\left(\bm{P}_1^{-1} \bm{Q}_2 \right)$, $K_0\left( s \right)$ in
\dref{eq:general:mmse:high_snr:K02} becomes
\begin{align}
K_0\left(s\right) &\simeq \frac{\Perm \left(\bm{Q}_2
\right)}{\left|\bm{P}_1\right| \Perm \left(\bm{P}_1^{-1} \bm{Q}_2
\right)} \sum_{i=1}^{N-1} \frac{\chi_i}{\vartheta_i +s},
\label{eq:general:mmse:high_snr:K03}
\end{align}
\noindent where
\begin{align}
\chi_i &= \frac{\zeta_{N-1}}{\prod_{k \neq i}^{N-1}
\left(\vartheta_k - \vartheta_i \right)}.
\label{eq:general:mmse:chii}
\end{align}
From \dref{eq:general:mmse:high_snr:ser1} and
\dref{eq:general:mmse:high_snr:K03}, we obtain
\begin{align}
P^{\infty}_s = \left(G_a \bar{\gamma}\right)^{-G_d} +
o\left(\bar{\gamma}^{-G_d} \right),
\label{eq:general:mmse:high_snr:asymp}
\end{align}
\noindent where the diversity order and array gain, $G_d$ and $G_a$
respectively, are given by
\begin{align}
G_d &= n_R-N+1,
\end{align}
and
\begin{align}
G_a &= \frac{\Perm \left(\bm{Q}_2 \right)}{\left|\bm{P}_1\right|
\Perm \left(\bm{P}_1^{-1} \bm{Q}_2 \right) } \mathcal{I}\left(
\bm{P} \right), \label{eq:general:mmse:high_snr:array_gain}
\end{align}
where
\begin{align}
\mathcal{I} \left( \bm{P} \right) \!=\! \frac{1}{\pi g^L}
\sum_{i=1}^{N-1} \frac{\chi_i}{\vartheta_i} \int_0^T  \frac{  \left(
\sin^2 \theta \right)^{L+1}}{\frac{g}{\vartheta_i} + \sin^2 \theta }
d\theta. \label{eq:general:mmse:high_snr:integral1}
\end{align}
The integrals in \dref{eq:general:mmse:high_snr:integral1} can be
solved in closed form as in \cite{Dush_mmse_dual_jnl}. Hence, the
final result becomes
\begin{align}
\mathcal{I} \left( \bm{P} \right) \!=\! \frac{1}{g^L}
\sum_{i=1}^{N-1} \frac{\chi_i}{\vartheta_i} J_{L+1}
\left(T,\frac{g}{\vartheta_i}\right),
\label{eq:general:mmse:high_snr:integral2}
\end{align}
\noindent where
\begin{align}
J_{m} \left(c,a\right) &= \frac{1}{\pi}\int_0^c
\frac{\sin^{2m}\theta}{a + \sin^2\theta} d\theta.
\label{eq:general:mmse:jm}
\end{align}
\section{Simulations and Numerical Results}\label{app:general:results}
\begin{figure}[!h]
\centering
\includegraphics[scale=0.65]{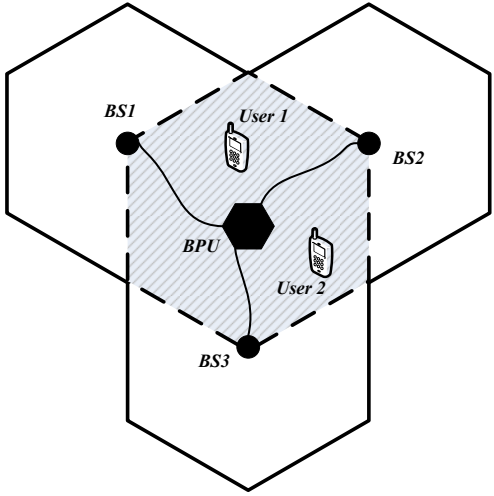}
\noindent \caption{Network MIMO/edge-excited cell scenario where
three base stations serve users in a three-sector cluster. To reduce
clutter, only two users are shown.} \label{fig:general:NW-MIMO_sim}
\end{figure}

\begin{figure}[h]
\centerline{\includegraphics*[scale=0.65]{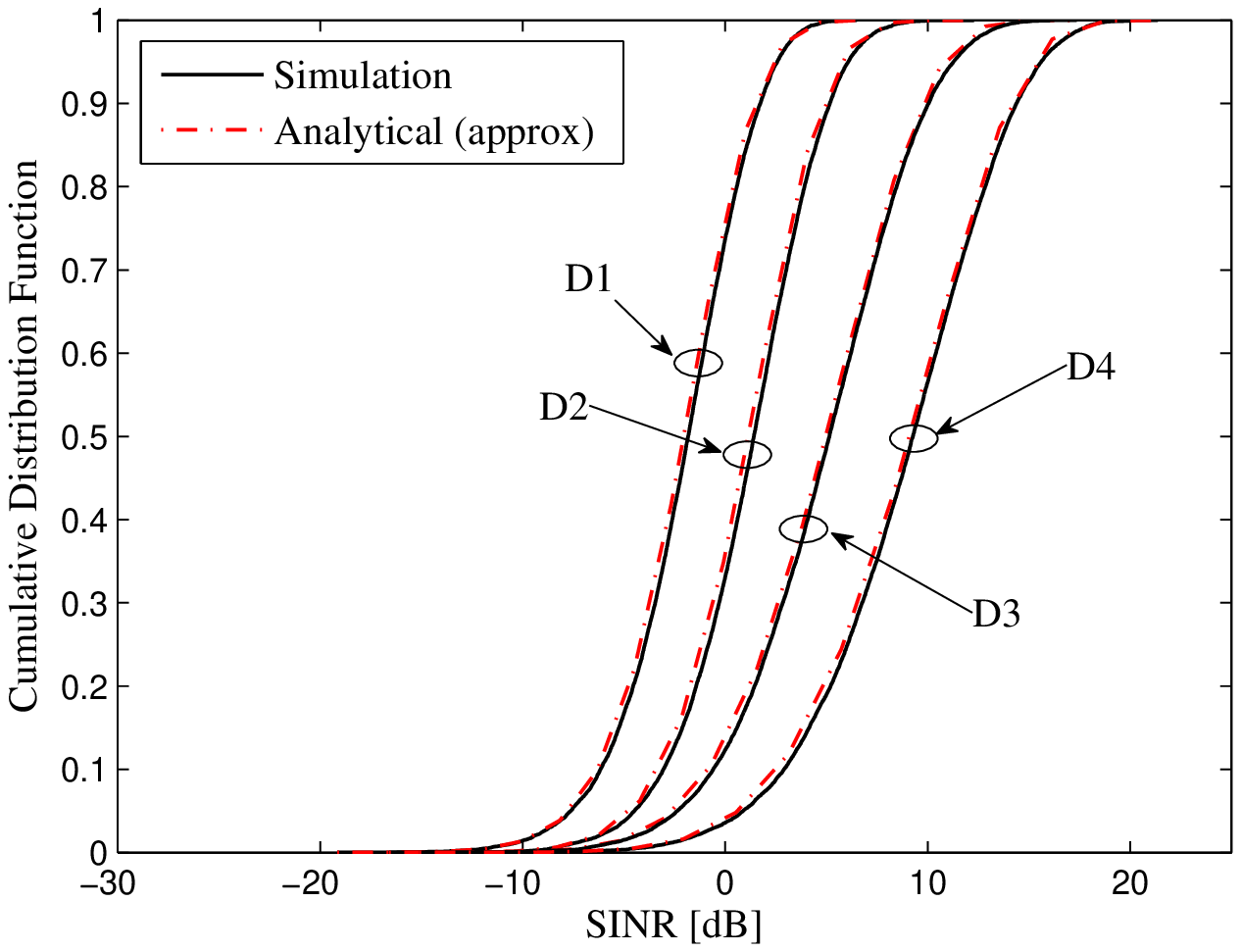}}
\caption{Approximate and simulated SINR CDF results for the $N=3$,
$n_R=3$ scenario. Results are shown for the first of three users for
four arbitrary drops and a MMSE receiver.} \label{fig:general:fig4}
\end{figure}

\begin{figure}[h]
\centerline{\includegraphics*[scale=0.65]{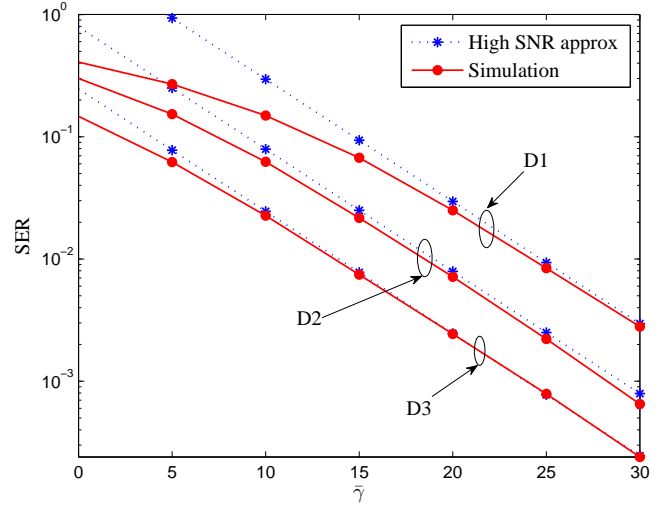}}
\caption{Approximate and simulated SER results for the $N=3$,
$n_R=3$ scenario with QPSK modulation. Results are shown for the
first of the three users for three arbitrary drops and a MMSE
receiver.} \label{fig:general:fig5}
\end{figure}

\begin{figure}[h]
\centerline{\includegraphics*[scale=0.65]{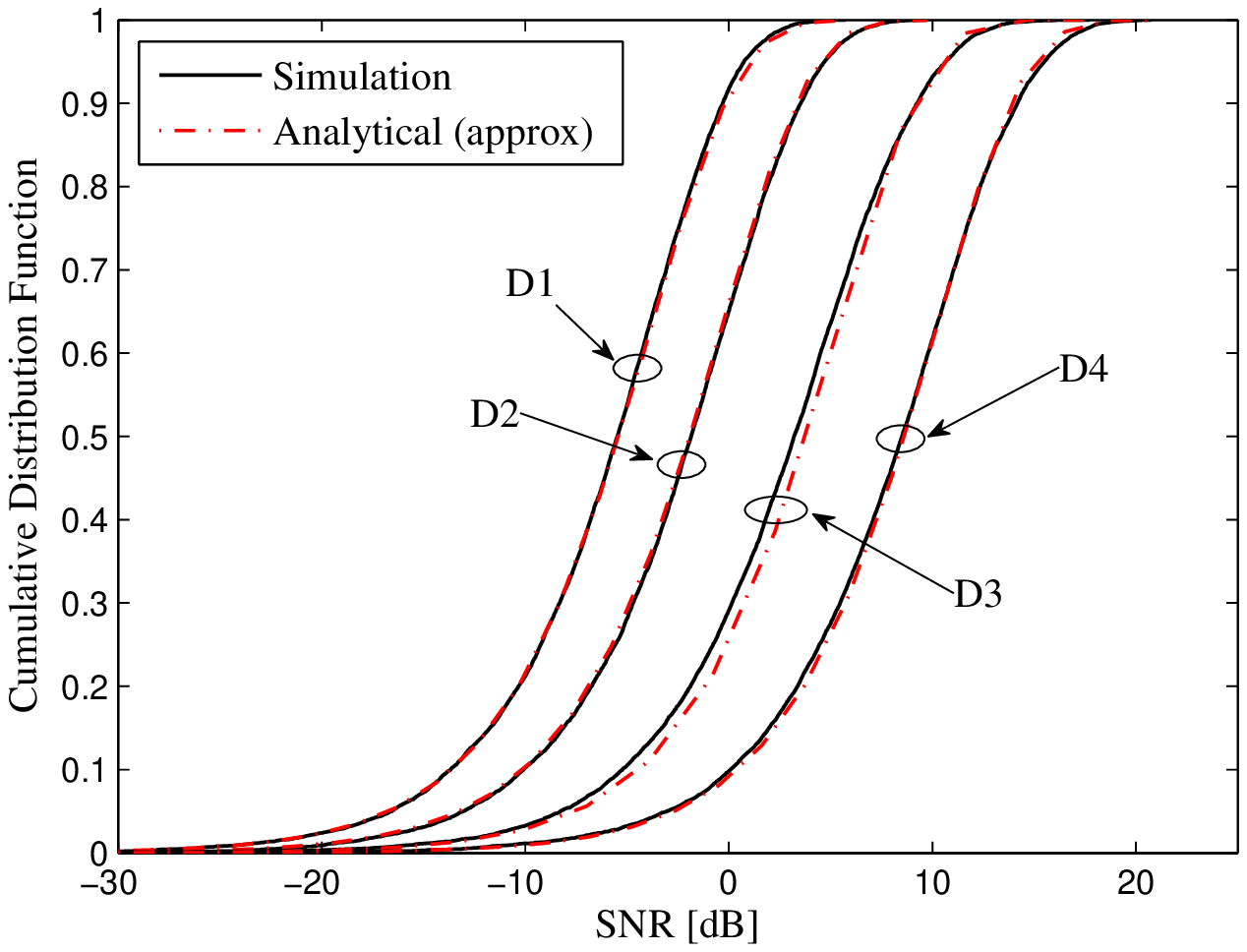}}
\caption{Approximate and simulated SNR CDF results for the $N=3$,
$n_R=3$ scenario. Results are shown for the first of three users for
four arbitrary drops and a ZF receiver.} \label{fig:general:fig2}
\end{figure}

\begin{figure}[h]
\centerline{\includegraphics*[scale=0.65]{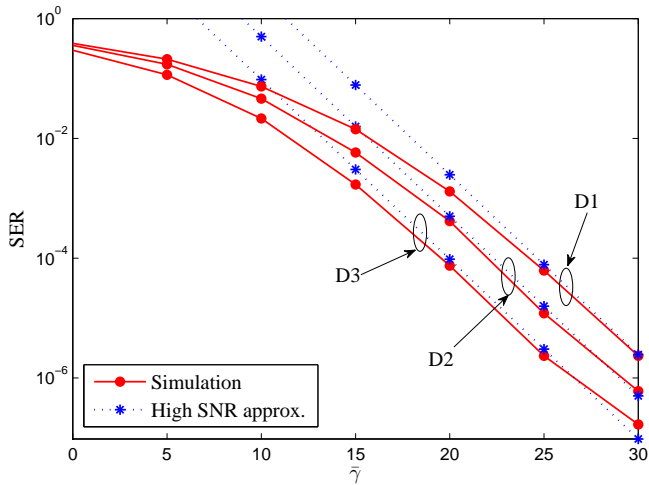}}
\caption{Approximate and simulated SER results for $N=4$, $n_R=6$,
i.e., two receive antennas at each BS with QPSK modulation. Results
are shown for the first of four users for three arbitrary drops and
a MMSE receiver.} \label{fig:general:fig6}
\end{figure}

\begin{figure}[h]
\centerline{\includegraphics*[scale=0.65]{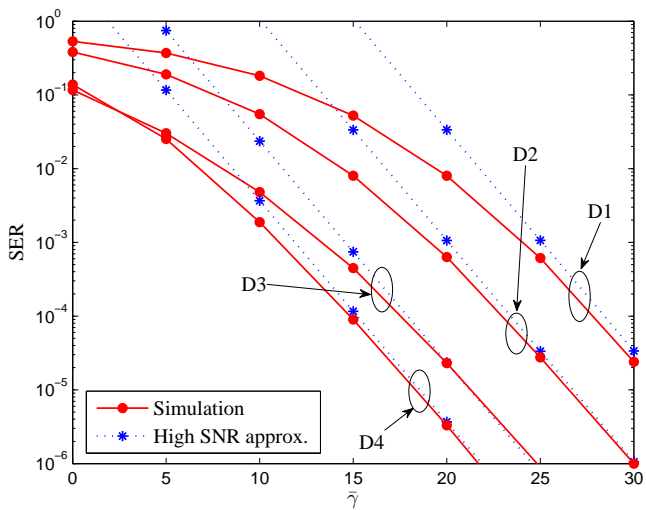}}
\caption{Approximate and simulated SER results for the $N=4$,
$n_R=6$, i.e., two receive antenna at each BS scenario with QPSK
modulation. Results are shown for the first of four users for four
arbitrary drops and a ZF receiver.} \label{fig:general:fig3}
\end{figure}

\begin{figure}[h]
\centerline{\includegraphics*[scale=0.65]{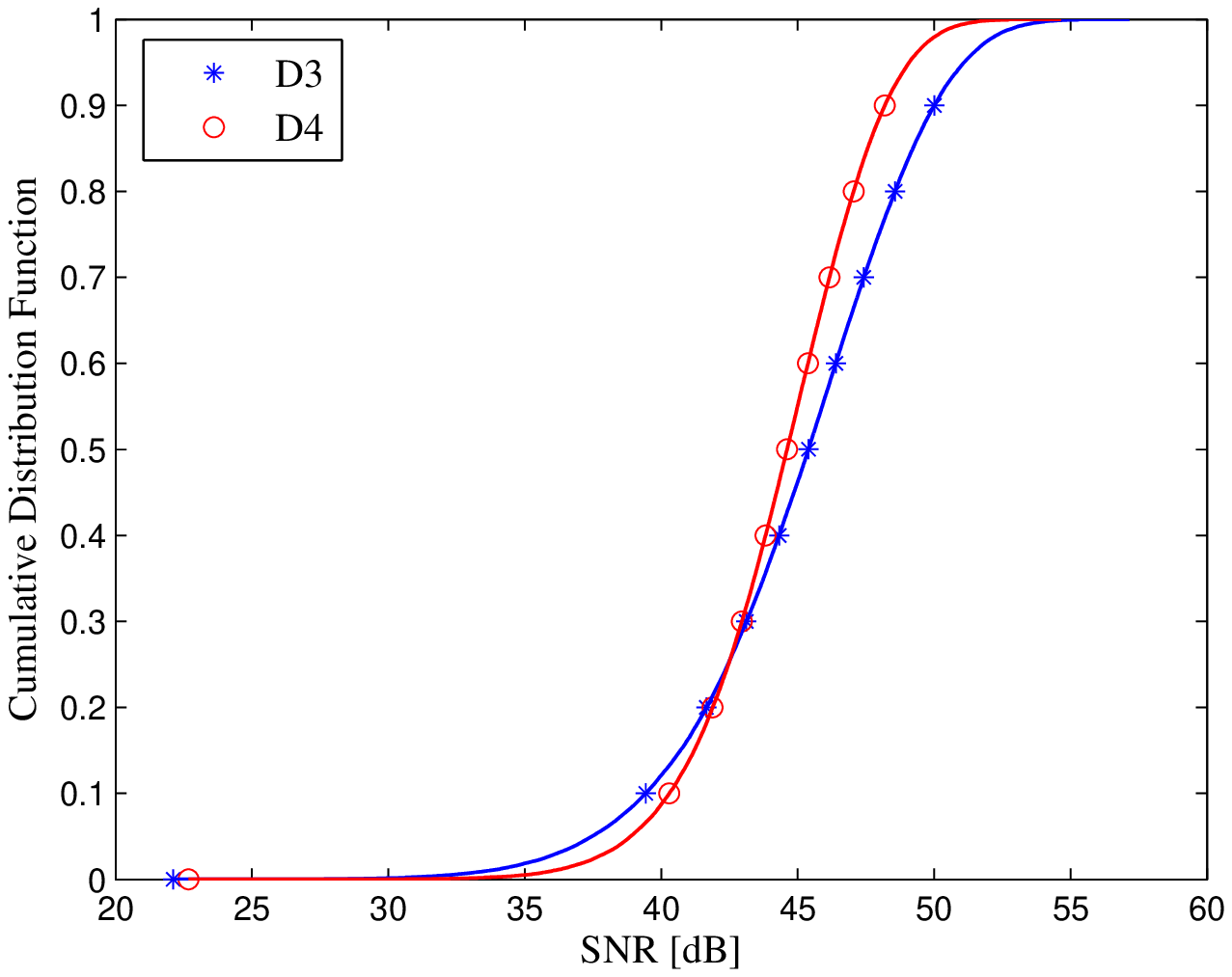}}
\caption{Simulated SNR CDF results for the $N=4$, $n_R=6$, i.e., two
receive antenna at each BS scenario. Results are shown for the first
of four users for D3 and D4 drops in Fig. \ref{fig:general:fig3} and
a ZF receiver.} \label{fig:general:fig8}
\end{figure}

In this section, we simulate the macrodiversity system shown in Fig.
\ref{fig:general:NW-MIMO_sim}, where three base stations (BSs)
collaborate via a central backhaul processing (BPU) in the shaded
three sector cluster. This simulation environment was also used in
\cite{Dush_mmse_dual_conf} and is sometimes referred to as an
edge-excited cell. We consider the three BS scenario having either a
single antenna or two antennas each to give $n_R=3$ or $n_R=6$
respectively. In the shaded coverage area of this edge-excited cell,
we drop three or four users uniformly in space giving $N=3$ or
$N=4$. For each user, lognormal shadow fading and path loss is
considered, where the standard deviation of the shadowing is 8dB and
the path loss exponent is $\gamma=3.5$. The transmit power of the
sources is scaled so that the best signal received at the three BS
locations is greater than 3dB at least $95\%$ of the time. Even
though the analysis in this paper is valid for any set of channel
powers, the above methodology allows us to investigate the accuracy
of the performance matrices for realistic sets of channel powers.\\
In Figs. \ref{fig:general:fig4}, \ref{fig:general:fig5} and
\ref{fig:general:fig2}, the case of three single antenna users and
three distributed BSs with a single receiver antenna is considered.
Here, we investigate both the approximate SINR distributions and the
approximate SER results for an MMSE receiver. In Fig.
\ref{fig:general:fig4}, the approximate CDFs of the output SINR are
plotted alongside the simulated CDFs. Results are shown for four
random drops and, the results are for a particular user (the first
of the three). The agreement between the CDFs is shown to be
excellent. Note that this agreement is good across all drops, from
D1 which has a very poor SINR performance to D4 with a much higher
SINR performance. The use of physically motivated drops rather than
ad-hoc scenarios is useful as it assesses the accuracy of the
analysis in plausible channel conditions.\\
In Fig. \ref{fig:general:fig5}, the approximate SER curve is plotted
alongside the simulated values. Results are shown for three drops
and QPSK modulation. The agreement between the SER results is shown
to be excellent across all three drops at SERs below $10^{-2}$.
Again, this agreement is observed over a wide range with D1 having
much higher SERs than D3. In Fig. \ref{fig:general:fig5} and also in
Figs. \ref{fig:general:fig6}-\ref{fig:general:fig3} the SER is
plotted against the transmit SNR, $\bar{\gamma}$. This is chosen
instead of the receive SNR to separate the curves so that the drops
are visible and are not all superimposed, which tends to happen when
SER is plotted against receive SNR.\\
In Fig. \ref{fig:general:fig2}, the approximate CDFs of the SNR are
plotted alongside the simulated CDFs for a ZF receiver. Results are
shown for four random drops. This is the companion plot to Fig.
\ref{fig:general:fig4} with the same system but a ZF receiver rather
than an MMSE receiver. The accuracy of the results in Fig.
\ref{fig:general:fig4} and Fig. \ref{fig:general:fig2} is
interesting, especially when you observe that the Fig.
\ref{fig:general:fig4} analysis uses \dref{eq:general:mmse:CDF2}, a
simple mixture of 3 exponentials, and Fig. \ref{fig:general:fig2}
uses \dref{eq:general:zf:CDF2} which is a single exponential in this
case.\\
In Fig. \ref{fig:general:fig6} and \ref{fig:general:fig3}, the case
of four single antenna users and six distributed receive antennas
(two at each BS location) is considered. High SNR SER curves are
plotted alongside the simulated values. Results are shown for both
MMSE (Fig. \ref{fig:general:fig6}) and ZF (Fig.
\ref{fig:general:fig3}) with QPSK modulation. The agreement between
the simulated SER and the high SNR approximation is shown to be less
accurate than in Fig. \ref{fig:general:fig5}, with very close
agreement requiring low error rates around $10^{-4}$. This is
unsurprising, as the greater number of system dimensions gives
greater freedom for the channel powers to vary substantially over
the links.\\
The results in Fig. \ref{fig:general:fig3} are very informative
concerning macrodiversity combining and highlight the difficulties
in predicting performance from the $\bm{P}$ matrix. Consider the
simple SIR metric given by the sum of the first column of $\bm{P}$
(the total long term received power from the desired user 1) divided
by the sum of columns 2,3 and 4 (the total long term interfering
power). In drops D1, D2 and D3 the SIR is -19dB, -2.5dB and 6.5dB.
As the SIR increases, the SER in Fig. \ref{fig:general:fig3} drops.
This is also shown by the $\tilde{K}_0$ metric in
\dref{eq:general:zf:high_snr:K0:2} which gives 17000, 323 and 13 for
drops D1, D2 and D3. As SER increases with $\tilde{K}_0$ both the
$\tilde{K}_0$ metric and the simple SIR metric give the same
performance ranking with D3 the best and D1 the worst. The fourth
drop, D4, is the interesting case. Here, the SIR is -10dB, which is
lower than both D2 and D3. Hence, from Fig. \ref{fig:general:fig3}
D4 has a better SER performance than D2 and D3 despite having a
worse SIR. In order to understand this, consider the $\bm{P}$ matrix
for drop D4,
\begin{align}
\bm{P}_{D4} &=  \left(
\begin{array}{cccc}
0.2061 & 1.3941  & 1.1034 & 4.6938  \\
0.2061 & 1.3941  & 1.1034 & 4.6938 \\
2.2923 & 16.8146  & 0.0857 & 0.6790  \\
2.2923 & 16.8146  & 0.0857 & 0.6790  \\
0.8361 & 3.4834  & 2.8181 & 0.6700  \\
0.8361 & 3.4834  & 2.8181 & 0.6700  \\
\end{array}  \right).
\label{eq:general:P_D4}
\end{align}
Again, it is difficult to see why D4 performs the best, since the
strongest long term power is on antennas $3 \& 4$ which also have
the strongest interference from user 2. Performance is clearly a
complex issue and is strongly related to the noise inflation caused
by the inverse operation in ZF reception (see
\dref{eq:general:zf:snr}). To verify this behavior, Fig.
\ref{fig:general:fig8} presents SNR curves for user 1 for drops D3
and D4 at $\bar{\gamma}=20$dB. As can be seen, the lower tail for D3
is higher than for D4 and this explains the increased SER. Although
a simple exploration of the $\bm{P}$ matrix makes it difficult to
predict D4 as the highest performing drop, the $\tilde{K}_0$ metric
captures this behavior as $\tilde{K}_0=1.3$, the lowest value for
all drops. In summary, the performance as a function of $\bm{P}$ is
difficult to predict without the analytical tools provided and here
the metric $\tilde{K}_0$ is particularly useful.
\section{Conclusion}\label{app:general:conclusion}
The performance of MMSE and ZF receivers is well-known in
microdiversity systems where the receive antennas are co-located.
However, in the macrodiversity case, closed form performance
analysis is a long-standing, unsolved research problem. In this
paper, we make the progress towards solving this problem for the
general case of an arbitrary number of transmit and receive
antennas. The analysis is based on a derivation which targets the
characteristic function of the output SINR. This leads to an
expected value which is highly complex in its exact form, but can be
simplified by the use of an extended Laplace type approximation. The
SINR distribution is shown to have a remarkably simple form as a
generalized mixture of exponentials. Also, the asymptotic SER
results produce a remarkably compact metric which captures a large
part of the functional relationship between the macrodiversity power
profile and SER.
\appendices
\section{Calculation of $\left|\bm{D}\right|\Perm \left(\bm{D}^{-1} \bm{Q}_2 \right)$}\label{app:general:zf_denom_expect}
The permanent of the denominator in \dref{eq:general:zf:CF6} can be
expanded as
\begin{align}
\Perm \left(\bm{D}^{-1} \bm{Q}_2 \right) &= \sum_\sigma \perm \left(
\left( \bm{D}^{-1}\bm{Q}_2 \right)_{\sigma_{N-1,n_R}}^{\{N-1\}}
\right), \label{eq:general:zf:CF7}
\end{align}
where $\sigma_{N-1,n_R}$ is an ordered subset of $\{n_R\}=\{1,2,
\dots,n_R \}$ of length $N-1$ and the sum is over all $\binom
{n_R}{N-1}$ such subsets. Noting the fact that $\perm
\left(\bm{\Sigma}\bm{X} \right)=\left|\bm{\Sigma}\right|
\perm\left(\bm{X}\right)$, for a square diagonal matrix
$\bm{\Sigma}$ and \dref{eq:general:axiom2}, \dref{eq:general:zf:CF7}
can be further simplified to give
\begin{align}
\Perm \left(\bm{D}^{-1} \bm{Q}_2 \right) &= \sum_\sigma \frac{\perm
\left( \left( \bm{Q}_2 \right)_{\sigma_{N-1,n_R}}^{\{N-1\}}
\right)}{\left|\bm{D}_{\sigma_{N-1,n_R}}^{\{N-1\}} \right|}.
\label{eq:general:zf:CF8}
\end{align}
Using \dref{eq:general:zf:CF8}, the denominator in
\dref{eq:general:zf:CF6} becomes
\begin{align}
\left|\bm{D}\right| \Perm \left(\bm{D}^{-1} \bm{Q}_2 \right) =
\sum_{\sigma}
&\left|\bm{D}_{\bar{\sigma}_{L,n_R}} \right| \nonumber \\
\times & \perm \left( \left( \bm{Q}_2
\right)_{\sigma_{N-1,n_R}}^{\{N-1\}} \right),
\label{eq:general:zf:CF9}
\end{align}
where $\bar{\sigma}_{L,n_R}$ is the ordered subset of length $L$ of
$\{1,2, \dots,n_R \}$ which does not belong to $\sigma_{N-1,n_R}$
and  $L=n_R-N+1$. Expanding $\left|\bm{D}_{\bar{\sigma}_{L,n_R}}
\right|$ gives
%
%
%
%
\begin{align}
\left|\bm{D}_{\bar{\sigma}_{L,n_R}} \right| &= \sum_{i=0}^{L}
\left(\frac{-jt}{\sigma^2}\right)^i \Tr_i \left(
\left(\bm{P}_1\right)_{\bar{\sigma}_{L,n_R}} \right).
\label{eq:general:zf:denominator:expectation1}
\end{align}
Substituting \dref{eq:general:zf:denominator:expectation1} in
\dref{eq:general:zf:CF9} gives the desired result
\begin{align}
\left|\bm{D}\right| \Perm \left(\bm{D}^{-1} \bm{Q}_2 \right) =
\sum_{i=0}^{L} \left(-jt\right)^i \tilde{\varphi}_i,
\label{eq:general:zf:denominator:expectation2_app}
\end{align}
\noindent where
\begin{align}
\tilde{\varphi}_i &= \sum_{\sigma} \Tr_i \left(
\left(\bm{P}_1\right)_{\bar{\sigma}_{L,n_R}} \right) \perm \left(
\left( \bm{Q}_2 \right)_{\sigma_{N-1,n_R}}^{\{N-1\}} \right)
\left(\sigma^2 \right)^{-i}.
\label{eq:general:zf:denominator:expectation:varphi_app}
\end{align}
\section{Calculation of $E \left\{ \left|\sigma^2 \bm{I} + \bm{H}_2^H \bm{H}_2\right|
\right \}$}\label{app:general:mmse_expectation1}
Similar expectation results for random determinants can also be
found in \cite{Dush_sum_cap_jnl}. However, for completeness we
present the particular result needed for the MMSE analysis here. Let
$\lambda_1, \lambda_2, \dots , \lambda_{N-1}$ be the ordered
eigenvalues of $\bm{H}_2^H \bm{H}_2$. Since $n_R \geq N$, all
eigenvalues are non zero. Then
\begin{align}
E \left\{ \left|\sigma^2 \bm{I} + \bm{H}_2^H \bm{H}_2\right| \right
\} &= E \left\{ \prod_{i=1}^{N-1} \left( \sigma^2 + \lambda_i
\right) \right
\} \\
\!&=\! E \!\left\{ \sum_{i=0}^{N-1} \Tr_{i} \!\left( \bm{H}_2^H
\bm{H}_2 \right)\left(\sigma^2 \right)^{N-i-1} \!\! \right \},
\label{eq:general:mmse_expectation:1}
\end{align}
where \dref{eq:general:mmse_expectation:1} is from
\cite[1.2.9]{Horn85} and \cite[1.2.12]{Horn85}. Therefore, the
building block of this expectation is $E \left\{\Tr_{i} \left(
\bm{H}_2^H \bm{H}_2 \right) \right \}$. From \cite[1.2.12]{Horn85},
\begin{align}
\Tr_{i} \left( \bm{H}_2^H \bm{H}_2 \right) &= \sum_{\sigma} \left|
\left( \bm{H}_2^H \bm{H}_2 \right)_{\sigma_{i,N-1}} \right|.
\label{eq:general:mmse_expectation:2}
\end{align}
Therefore, from Lemma
\ref{lemma:general:expected_square_determinant}
\begin{align}
E \left\{ \Tr_{i} \left( \bm{H}_2^H \bm{H}_2 \right) \right \} &=
\sum_{\sigma} \Perm \left( \left(\bm{Q}_2 \right)^{\sigma_{i,N-1}}
\right), \label{eq:general:mmse_expectation:3}
\end{align}
\noindent where the $n_R \times \left( N-1 \right)$ matrix,
$\bm{Q}_2$, is given by
\begin{align}
E \left\{\bm{H}_2 \circ \bm{H}_2 \right\} &= \bm{Q}_2.
\end{align}
Then, the final expression becomes
\begin{align}
E \left\{ \left|\sigma^2 \bm{I} + \bm{H}_2^H \bm{H}_2\right| \right
\} \!&=\!\! \sum_{i=0}^{N-1}\!\! \sum_{\sigma} \Perm \left(
\left(\bm{Q}_2 \right)^{\sigma_{i,N-1}} \right) \left(\sigma^2
\right)^{N-i-1}.
\label{eq:general:mmse_expectation:final_expression}
\end{align}
\section{Calculation of $\left|\bm{D}\right| E \left\{ \left|\sigma^2 \bm{I} + \bm{H}_2^H
\bm{D}^{-1} \bm{H}_2\right| \right
\}$}\label{app:general:mmse_expectation2_app}
As a simple extension of expectation in the numerator of
\dref{eq:general:mmse:CF6}, the expectation in the denominator can
be calculated using
\dref{eq:general:mmse_expectation:final_expression} to give
\begin{align}
E \left\{ \left|\sigma^2 \bm{I} + \bm{H}_2^H \bm{D}^{-1}
\bm{H}_2\right| \right \} \!\! &=\!\! \sum_{k=0}^{N-1} \psi_k
\left(-jt\right) \left(\sigma^2 \right)^{N-k-1},
\label{eq:general:mmse_denominator:expectation}
\end{align}
\noindent where
\begin{align}
\psi_k \left(-jt\right) &= \sum_{\sigma} \Perm \left(
\left(\bm{D}^{-1} \bm{Q}_2 \right)^{\sigma_{k,N-1}} \right),
\label{eq:general:mmse_denominator:expectation:psi1}
\end{align}
and
\begin{align}
\psi_0 \left(-jt\right) &= 1.
\label{eq:general:mmse_denominator:expectation:psi_zero}
\end{align}
The term in \dref{eq:general:mmse_denominator:expectation:psi1} can
be simplified using \dref{eq:general:axiom2} and Corollary
\ref{corollary:general:expected_diagonal_rectangular_determinant} to
obtain
\begin{align}
\psi_k \left(-jt\right) &=  \sum_\sigma \frac{\Perm \left( \left(
\bm{Q}_2
\right)_{\sigma_{k,n_R}}^{\{N-1\}}\right)}{\left|\bm{D}_{\sigma_{k,n_R}}\right|}.
\label{eq:general:mmse_denominator:expectation:psi2}
\end{align}
Then,
\begin{align}
\left|\bm{D}\right| E \left\{ \left|\sigma^2 \bm{I} + \bm{H}_2^H
\bm{D}^{-1} \bm{H}_2\right| \right \} \!\! &=\!\! \sum_{k=0}^{N-1}
\xi_k \left(-jt\right) \left(\sigma^2 \right)^{N-k-1},
\label{eq:general:mmse_denominator:expectation}
\end{align}
\noindent where $\xi_k\left(-jt\right) =\left|\bm{D}\right|
\psi_k\left(-jt\right)$. From
\dref{eq:general:mmse_denominator:expectation:psi2}, we can get
\begin{align}
\xi_k\left(-jt\right) &= \sum_\sigma
\left|\bm{D}_{\bar{\sigma}_{n_R-k,n_R}} \right| \Perm \left( \left(
\bm{Q}_2 \right)_{\sigma_{k,n_R}}^{\{N-1\}}\right),
\label{eq:general:mmse_denominator:expectation:xi1}
\end{align}
\noindent where $\bar{\sigma}_{n_R-k,n_R}$ is a length $n_R -k$
subset of $\left\{ 1, \dots, n_R \right\}$ which does not belong to
$\sigma_{k,n_R}$. Therefore, it is apparent that
$\xi_k\left(-jt\right)$ is a polynomial of degree $n_R-k$. It is
clear from \dref{eq:general:mmse_denominator:expectation} that, when
$\sigma^2=0$, \dref{eq:general:mmse_denominator:expectation}
collapses to \dref{eq:general:zf:CF9}. Clearly, $\left|\bm{D}\right|
E \left\{ \left|\sigma^2 \bm{I} + \bm{H}_2^H \bm{D}^{-1}
\bm{H}_2\right| \right \}$ is a polynomial of degree $n_R$, as
$\xi_0\left(-jt\right)=\left|\bm{D}\right|$ is the highest degree
polynomial term in \dref{eq:general:mmse_denominator:expectation}.
Then,
\begin{align}
\left|\bm{D}_{\bar{\sigma}_{n_R-k,n_R}} \right| &=
\sum_{i=0}^{n_R-k} \left(\frac{-jt}{\sigma^2}\right)^i \Tr_i \left(
\left(\bm{P}_1\right)_{\bar{\sigma}_{n_R-k,n_R}} \right).
\label{eq:general:mmse_denominator:expectation:anonymous1}
\end{align}
Hence
\begin{align}
\xi_k\left(-jt\right) = \sum_\sigma \sum_{i=0}^{n_R-k}
\left(\frac{-jt}{\sigma^2}\right)^i &\Tr_i \left(
\left(\bm{P}_1\right)_{\bar{\sigma}_{n_R-k,n_R}} \right) \nonumber \\
&\times \Perm \left( \left( \bm{Q}_2
\right)_{\sigma_{k,n_R}}^{\{N-1\}}\right),
\label{eq:general:mmse_denominator:expectation:xi2}
\end{align}
so that $\xi_k\left(-jt\right)$ becomes
\begin{align}
\xi_k\left(-jt\right) &= \sum_{i=0}^{n_R-k}
\left(\frac{-jt}{\sigma^2}\right)^i \hat{\varphi}_{ik} \\
&= \sum_{i=0}^{n_R} \left(\frac{-jt}{\sigma^2}\right)^i
\hat{\varphi}_{ik},
\label{eq:general:mmse_denominator:expectation:xi3}
\end{align}
\noindent where
\begin{align}
\hat{\varphi}_{ik} &= \left. \sum_\sigma \Tr_i \left(
\left(\bm{P}_1\right)_{\bar{\sigma}_{n_R-k,n_R}} \right) \Perm
\left( \left( \bm{Q}_2 \right)_{\sigma_{k,n_R}}^{\{N-1\}}\right)
\right.,
\label{eq:general:mmse_denominator:expectation:acute_varphi1}
\end{align}
and $\hat{\varphi}_{i0}$ simplifies to give
\begin{align}
\hat{\varphi}_{i0} &=  \Tr_i \left(\bm{P}_1 \right).
\label{eq:general:mmse_denominator:expectation:acute_varphi_i_zero}
\end{align}
Equation \dref{eq:general:mmse_denominator:expectation:xi3} follows
from the fact that
\begin{align}
\Tr_i \left( \left(\bm{P}_1\right)_{\bar{\sigma}_{n_R-k,R}} \right)
=0 \quad \mbox{for}  \quad i > n_R-k.
\end{align}

Therefore, \dref{eq:general:mmse_denominator:expectation} can be
written as
\begin{align}
\left|\bm{D}\right| E \left\{ \left|\sigma^2 \bm{I} + \bm{H}_2^H
\bm{D}^{-1} \bm{H}_2\right| \right \} = \sum_{k=0}^{N-1}
&\sum_{i=0}^{n_R} \left(-jt\right)^i \hat{\varphi}_{ik}   \nonumber \\
&\times \left(\sigma^2 \right)^{N-i-k-1},
\end{align}
\noindent which is in turn can be given as
\begin{align}
\left|\bm{D}\right| E \left\{ \left|\sigma^2 \bm{I} + \bm{H}_2^H
\bm{D}^{-1} \bm{H}_2\right| \right \} &= \sum_{i=0}^{n_R}
\left(-jt\right)^{i} \varphi_i,
\label{eq:general:mmse_denominator:expectation1_app}
\end{align}
\noindent where
\begin{align}
\varphi_i &=  \sum_{k=0}^{N-1} \hat{\varphi}_{ik} \left(\sigma^2
\right)^{N-i-k-1}.
\label{eq:general:mmse_denominator:expectation:varphi1_app}
\end{align}

\newpage
\begin{IEEEbiography}[{\includegraphics[width=1in,height=1.25in,clip,keepaspectratio]{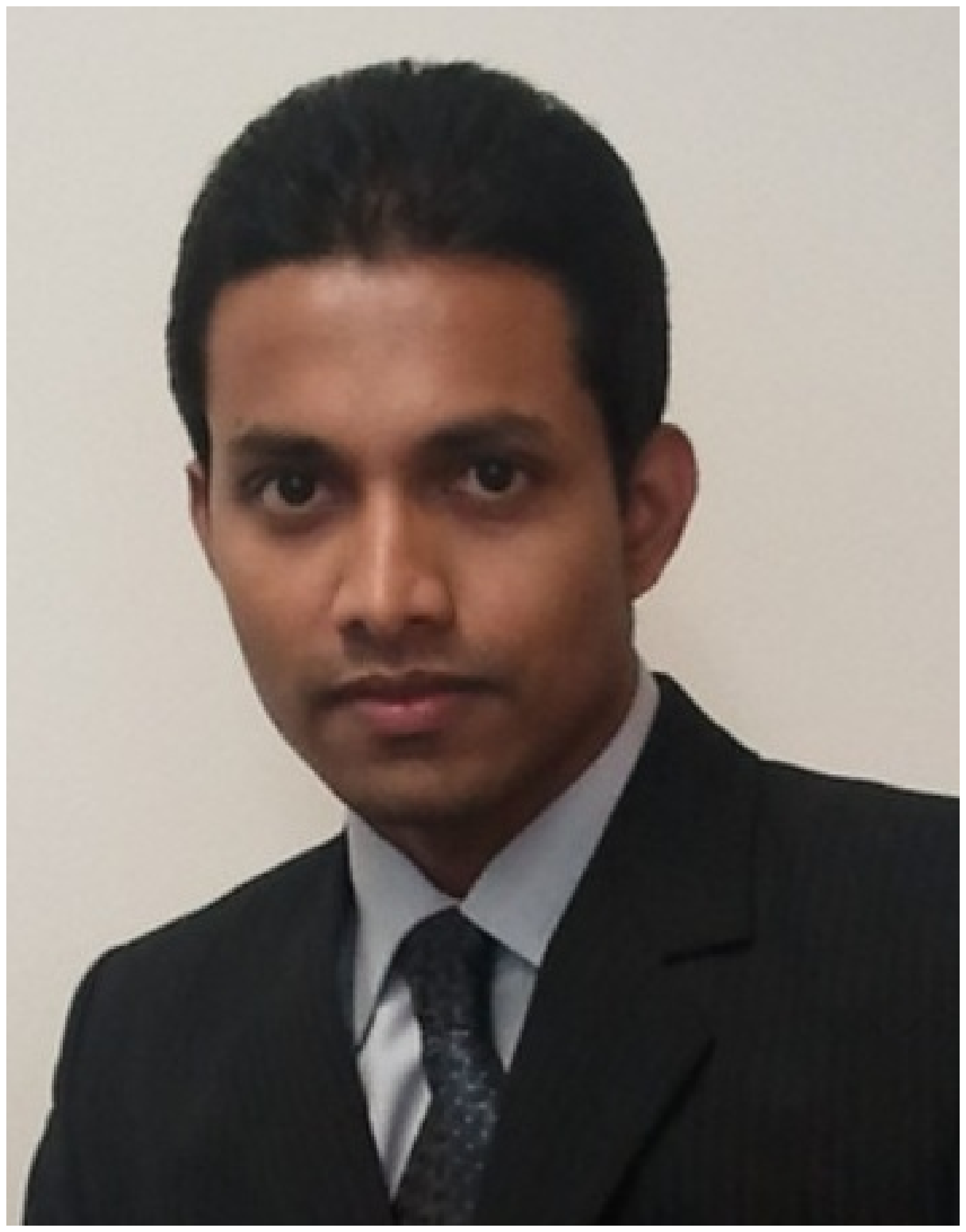}}]{Dushyantha Basnayaka}
(S'11-M'12) was born in 1982 in Colombo, Sri Lanka. He received the
B.Sc.Eng degree with 1\textsuperscript{st} class honors from the
University of Peradeniya, Sri Lanka, in Jan 2006. He is currently
working towards for his PhD degree in Electrical and Computer
Engineering at the University of Canterbury, Christchurch, New
Zealand.\\
He was an instructor in the Department of Electrical and Electronics
Engineering at the University of Peradeniya from Jan 2006 to May
2006. He was a system engineer at MillenniumIT (a member company of
London Stock Exchange group) from May 2006 to Jun. 2009. Since Jul.
2009 he is with the communication research group at the University
of Canterbury, New Zealand.\\
D. A. Basnayaka is a recipient of University of Canterbury
International Doctoral Scholarship for his doctoral studies at UC.
His current research interest includes all the areas of digital
communication, especially macrodiversity wireless systems. He holds
one pending US patent as a result of his doctoral studies at UC.
\end{IEEEbiography}
\begin{IEEEbiography}{Peter Smith}
(M'93-SM'01) received the B.Sc degree in Mathematics and the Ph.D
degree in Statistics from the University of London, London, U.K., in
1983 and 1988, respectively. From 1983 to 1986 he was with the
Telecommunications Laboratories at GEC Hirst Research Centre. From
1988 to 2001 he was a lecturer in statistics at Victoria University,
Wellington, New Zealand. Since 2001 he has been a Senior Lecturer
and Associate Professor in Electrical and Computer Engineering at
the University of Canterbury in New Zealand. Currently, he is a full
Professor at the same department.\\
His research interests include the statistical aspects of design,
modeling and analysis for communication systems, especially antenna
arrays, MIMO, cognitive radio and relays.
\end{IEEEbiography}
\begin{IEEEbiography}{Philippa Martin}
(S'95-M'01-SM'06) received the B.E. (Hons. 1) and Ph.D. degrees in
electrical and electronic engineering from the University of
Canterbury, Christchurch, New Zealand, in 1997 and 2001,
respectively.  From 2001 to 2004, she was a postdoctoral fellow,
funded in part by the New Zealand Foundation for Research, Science
and Technology (FRST), in the Department of Electrical and Computer
Engineering at the University of Canterbury.  In 2002, she spent 5
months as a visiting researcher in the Department of Electrical
Engineering at the University of Hawaii at Manoa, Honolulu, Hawaii,
USA.  Since 2004 she has been working at the University of
Canterbury as a lecturer and then as a senior lecturer. Currently,
she is an Associate Professor at the same department. In 2007, she
was awarded the University of Canterbury, College of Engineering
young researcher award.  She served as an Editor for the IEEE
Transactions on Wireless Communications 2005-2008 and regularly
serves on technical program committees for IEEE conferences. \\
Her current research interests include multilevel coding, error
correction coding, iterative decoding and equalization, space-time
coding and detection, cognitive radio and cooperative communications
in particular for wireless communications
\end{IEEEbiography}

\end{document}